\documentclass[12pt]{iopart}

\usepackage{graphicx}
\usepackage{url}

\newcommand{\OO}[1]{{\mathcal O}(c^{-#1})}

\newcommand{\muas}[0]{\hbox{\rm $\mu$as}}

\newcommand{\ve}[1]{\mbox{\boldmath$#1$}}

\def\source{{\rm 0}}
\def\obs{{\rm 1}}

\let\oldbibitem\bibitem
\renewcommand\bibitem[2][]{\oldbibitem{#2}}

\begin{document}

\title[Numerical versus analytical accuracy of the light propagation formulas]
{Numerical versus analytical accuracy of the formulas for light propagation}

\author{Sergei A. Klioner, Sven Zschocke}

\address{Lohrmann Observatory, Dresden Technical University,
Mommsenstr. 13, 01062 Dresden, Germany}

\ead{Sergei.Klioner@tu-dresden.de}

\begin{abstract}

  Numerical integration of the differential equations of light
  propagation in the Schwarzschild metric shows that in some
  situations relevant for practical observations the
  well-known post-Newtonian solution for light propagation
  has an error up to 16 \muas. The aim of this work is to demonstrate this fact, 
  identify the reason for this error and to derive an analytical formula accurate
  at the level of 1 \muas\ as needed for high-accuracy astrometric
  projects (e.g., Gaia).

  An analytical post-post-Newtonian solution for the light propagation
  for both Cauchy and boundary problems is given for the Schwarzschild
  metric augmented by the PPN and post-linear parameters $\beta$,
  $\gamma$ and $\epsilon$. Using analytical upper estimates of each
  term we investigate which post-post-Newtonian terms may play a role
  for an observer in the solar system at the level of 1 \muas\ and
  conclude that only one post-post-Newtonian term remains important
  for this numerical accuracy. In this way, an analytical solution for
  the boundary problem for light propagation is derived. That solution
  contains terms of both post-Newtonian and post-post-Newtonian order,
  but is valid for the given numerical level of 1 \muas.  The derived
  analytical solution has been verified using the results of a
  high-accuracy numerical integration of differential equations of
  light propagation and found to be correct at the level well below 1
  \muas\ for arbitrary observer situated within the solar system.
  Furthermore, the origin of the post-post-Newtonian terms relevant
  for the microarcsecond accuracy is elucidated. We demonstrate that
  these terms result from an inadequate choice of the impact parameter
  in the standard post-Newtonian formulas. Introducing another impact
  parameter, that can be called ``coordinate-independent'', we
  demonstrate that all these terms disappear from the formulas.

\end{abstract}

\pacs{95.10.Jk, 95.10.Ce, 95.30.Sf, 04.25.Nx, 04.80.Cc}

%\submitto{\CQG}
%\maketitle

\section{Introduction}

It is well known that adequate relativistic modeling is indispensable
for the success of microarcsecond space astrometry. One of the most important
relativistic effects for astrometric observations in the solar system
is the gravitational light deflection. The largest contribution in the
light deflection comes from the spherically symmetric (Schwarzschild)
parts of the gravitational fields of each solar system body. Although
the planned astrometric satellites Gaia, SIM, etc. will not observe very
close to the Sun, they can observe very close to the giant planets also
producing significant light deflection. This poses the problem of
modeling this light deflection with a numerical accuracy of better
than 1 \muas.

The exact differential equations of motion for a light ray in the
Schwarzschild field can be solved numerically as well as analytically.
However, the exact analytical solution is given in terms of elliptic
integrals, implying numerical efforts comparable with direct numerical
integration, so that approximate analytical solutions are usually
used. In fact, the standard parametrized post-Newtonian (PPN) solution
is sufficient in many cases and has been widely applied. So far, there
was no doubt that the post-Newtonian order of approximation is
sufficient for astrometric missions even up to microarcsecond level of
accuracy, besides astrometric observations close to the edge of the
Sun. However, a direct comparison reveals a deviation between the
standard post-Newtonian approach and the high-accuracy numerical
solution of the geodetic equations. In particular, we have found a
difference of up to 16 \muas\ in light deflection for solar system
objects observed close to giant planets. This error has triggered
detailed numerical and analytical investigation of the problem.

Usually, in the framework of general relativity or the PPN formalism
analytical orders of smallness of various terms are considered. Here
the role of small parameter is played by $c^{-1}$, where $c$ is the
light velocity. Standard post-Newtonian and post-post-Newtonian
solutions are derived by retaining terms of relevant analytical orders
of magnitude. On the other hand, for practical calculations only
numerical magnitudes of various terms are relevant. In this work we
attempt to close this gap and combine the analytical
parametrized post-post-Newtonian solution with exact analytical
estimates of the numerical magnitudes of various terms. In
this way we derive a compact analytical solution 
for light propagation where all terms are indeed relevant at
the level of 1 \muas. The derived analytical solution is then verified
using high-accuracy numerical integration of the
differential equations of light propagation and found to be correct at
the level well below 1 \muas.

%%% notations

We use fairly standard notations: 

\begin{itemize}

\item $G$ is the Newtonian constant of gravitation.

\item $c$ is the velocity of light.

\item $\beta$ and $\gamma$ are the parameters of the Parametrized Post-Newtonian
(PPN) formalism which characterize possible deviation of the physical
reality from general relativity theory ($\beta=\gamma=1$ in general
relativity).

\item Lower case Latin indices $i$, $j$, \dots take values 1,
2, 3.

\item Lower case Greek indices $\mu$, $\nu$, \dots take values 0, 1,
2, 3.

\item Repeated indices imply the Einstein's summation irrespective of
their positions (e.g. $a^i\,b^i=a^1\,b^1+a^2\,b^2+a^3\,b^3$ and
$a^\alpha\,b^\alpha=a^0\,b^0+a^1\,b^1+a^2\,b^2+a^3\,b^3$).

\item A dot over any quantity designates the total derivative with
respect to the coordinate time of the corresponding reference system:
e.g. $\dot a=\displaystyle{da\over dt}$.

\item The 3-dimensional coordinate quantities (``3-vectors'') referred to
the spatial axes of the corresponding reference system are set in
boldface: $\ve{a}=a^i$.

\item The absolute value (Euclidean norm) of a ``3-vector'' $\ve{a}$ is
denoted as $|\ve{a}|$ or, simply, $a$ and can be computed as
$a=|\ve{a}|=(a^1\,a^1+a^2\,a^2+a^3\,a^3)^{1/2}$.

\item The scalar product of any two ``3-vectors'' $\ve{a}$ and $\ve{b}$
with respect to the Euclidean metric $\delta_{ij}$ is denoted by
$\ve{a}\,\cdot\,\ve{b}$ and can be computed as
$\ve{a}\,\cdot\,\ve{b}=\delta_{ij}\,a^i\,b^j=a^i\,b^i$.

\item The vector product of any two ``3-vectors'' $\ve{a}$ and $\ve{b}$
is designated by $\ve{a}\times\ve{b}$ and can be computed as
$\left(\ve{a}\times\ve{b}\right)^i=\varepsilon_{ijk}\,a^j\,b^k$, where
$\varepsilon_{ijk}=(i-j)(j-k)(k-i)/2$ is the fully antisymmetric
Levi-Civita symbol.

\item For any two vectors $\ve{a}$ and $\ve{b}$, the angle between
  them is designated as $\delta(\ve{a},\ve{b})$. Clearly, for an angle
  between two vectors one has $0\le\delta(\ve{a},\ve{b})\le\pi$. Angle
  $\delta(\ve{a},\ve{b})$ can be computed in many ways, for example,
  as $\delta
  (\ve{a},\ve{b})=\arccos\displaystyle{\ve{a}\cdot\ve{b}\over a\,b}$.

\end{itemize}

%% table of content
This paper is a concise exposition of the work performed in the
framework of the ESA project Gaia and published in a series of
preprints \cite{report1,report2,report3,report-proofs}.  The paper is
organized as follows. In Section~\ref{section-schwarzschild} we
present the exact differential equations for the light propagation in
the Schwarzschild field in harmonic gauge.  High-accuracy numerical
integrations of these equations are discussed in
Section~\ref{Section:numerical_integration}.  In
Section~\ref{section-standard-pN} we discuss the standard
post-Newtonian approximation and demonstrate its errors by direct
comparison with numerical results.  In
Section~\ref{section-ppN-solution} the analytical post-post-Newtonian
solution for the light propagation is given.
Section~\ref{section-boundary-problem} is devoted to the boundary
problem for the light propagation in post-post-Newtonian
approximation. Investigations of the post-post-Newtonian terms in the
formulas for the light deflection reveal that these terms can be
divided into two groups: ``regular'' (those which can be estimated as
${\rm const}\cdot {m^2\over d^2}$, where $m$ is the Schwarzschild
radius of the deflecting body and $\ve{d}$ is the impact parameter)
and ``enhanced'' (those which cannot be estimated like this and may
become substantially larger than the ``regular'' terms).  In
Section~\ref{section-physical-origin} we clarify the physical origin
of the ``enhanced'' post-post-Newtonian terms.  The results are
summarized in Section~\ref{section-conclusion}.

\section{Schwarzschild metric and null geodesics in harmonic coordinates}
\label{section-schwarzschild}

We need a tool to calculate the real numerical accuracy of some
analytical formulas for the light propagation in various
situations. To this end, we consider the exact Schwarzschild metric
and its null geodesics in harmonic gauge. Those exact differential
equations for the null geodesics will be solved numerically with high
accuracy (see below) and that numerical solution provides the required
reference.

\subsection{Metric tensor}

In harmonic gauge
\begin{equation}
\label{harmonic-conditions}
\frac{\partial \left( \sqrt{- g} \, g^{\alpha \beta} \right)}{\partial x^{\beta}}= 0
\end{equation}
\noindent
the components of the covariant metric tensor of the Schwarzschild
solution are given by
\begin{eqnarray}
g_{00} &=& - \frac{1-a}{1+a} ,
\nonumber\\
g_{0i} &=& 0 ,
\nonumber\\
g_{ij} &=& \left(1 + a \right)^2 \, \delta_{i j}  + 
\frac{a^2}{x^2} \, \frac{1+a}{1-a} \, x^i \, x^j \, ,
\label{exact_5}
\end{eqnarray}
\noindent
where
\begin{equation}
\label{a-def}
a={m\over x},
\end{equation}
\noindent
and $m=\frac{\displaystyle G \,M}{\displaystyle c^2}$ is the Schwarzschild
radius of a body with mass $M$. The contravariant components of the metric read
\begin{eqnarray}
g^{00} &=& \frac{1 + a}{1 - a} \,,
\nonumber\\
g^{0i} &=& 0 \,,
\nonumber\\
g^{ij} &=& \frac{1}{\left(1 + a \right)^2} \; \delta_{i j}
\; - \; \frac{a^2}{x^2} \; \frac{1}{\left(1 + a \right)^2}
\; x^i \, x^j \,.
\label{exact_10}
\end{eqnarray}
\noindent
Considering that the determinant of the metric can be computed as
\begin{equation}
  \label{eq:determinant}
  g=-(1+a)^4,
\end{equation}
\noindent
one can easily check that this metric satisfies the harmonic conditions
(\ref{harmonic-conditions}).

\subsection{Christoffel symbols}

The Christoffel symbols of second kind are defined as
\begin{eqnarray}
\Gamma^{\mu}_{\alpha \beta} &=& \frac{1}{2}\;g^{\mu \nu}\;
\left( \frac{\partial g_{\nu \alpha} }{\partial x^{\beta}} \;
+ \; \frac{\partial g_{\nu \beta} }{\partial x^{\alpha}} \; - \;
\frac{\partial g_{\alpha \beta} }{\partial x^{\nu}} \right) \,.
\label{christoffel}
\end{eqnarray}
\noindent
Using (\ref{exact_5}) and (\ref{exact_10}) one gets
\begin{eqnarray}
\Gamma^0_{0 i} &=& \frac{a}{x^2} \; \frac{1}{1 - a^2} \; x^i\,,
\nonumber\\
\Gamma^i_{0 0} &=& \frac{a}{x^2} \; \frac{1-a}{(1 + a)^3} \; x^i \,,
\nonumber\\
\Gamma^i_{j k} &=& \frac{a}{x^2} \; x^i \; \delta_{j k} \; - \;
\frac{a}{x^2} \; \frac{1}{1 + a} \; \left( x^j \; \delta_{i k} \; + \;
x^k \; \delta_{i j} \right)
\; - \; \frac{a^2}{x^4} \; \frac{2 - a}{1 - a^2}\;x^i\;x^j\;x^k\,.
\label{exact_15}
\end{eqnarray}
\noindent
All other Christoffel symbols vanish.

\subsection{Isotropic condition}
\label{section-isotropic-condition}

The conditions that a photon follows an isotropic geodesic can be
formulated as an equation for the four components of the coordinate
velocity $\dot x^{\alpha}$ of that photon:
\begin{eqnarray}
g_{\alpha \beta} \; \frac{d \, x^{\alpha}}{d \, \lambda} \;
\frac{d \, x^{\beta}}{d \, \lambda} &=& 0 \,,
\label{isotropic_5}
\end{eqnarray}
\noindent
$\lambda$ being the canonical parameter, or
\begin{equation}
\label{isotropic1}
g_{00}+{2\over c}\,g_{0i}\,\dot x^i+{1\over c^2}\,g_{ij}\,\dot x^i\,
\dot x^j=0 \, ,
\label{isotropic_10}
\end{equation}
\noindent
where $\dot x^i=dx^i/dt$ is the coordinate velocity of the photon.
Eq. (\ref{isotropic_10}) is a first integral of motion for the differential
equation for light propagation and must be valid for any point of an
isotropic geodesic. Substituting the ansatz $\dot{\ve{x}}=c\,s\,\ve{\mu}$,
where $\ve{\mu}$ is a unit coordinate direction of light propagation
($\ve{\mu}\cdot\ve{\mu}=1$) and $s=|\dot{\ve{x}}|/c$, into (\ref{isotropic_10})
one gets for metric (\ref{exact_5}):
\begin{eqnarray}
s &=& \frac{1 - a}{1 + a}\;
\left( 1 - a^2 + \frac{a^2}{x^2} (\ve{x} \cdot \ve{\mu})^2 \right)^{-1/2}\,.
\label{isotropic_15}
\end{eqnarray}
\noindent
This formula allows one to compute the absolute value of coordinate
velocity of light in the chosen reference system if the position of
the photon $x^i$ and the coordinate direction of its propagation
$\ve{\mu}$ are given.

\subsection{Equation of isotropic geodesics}

The geodetic equations
\begin{eqnarray}
\frac{d^2 x^{\mu}}{d \lambda^2} \; + \; \Gamma^{\mu}_{\alpha \beta} \;
\frac{d x^{\alpha}}{d \lambda}\,\frac{d x^{\beta}}{d \lambda} &=& 0
\label{geodetic_5}
\end{eqnarray}
\noindent
can be re-parametrized by coordinate time $t$ to give
\begin{eqnarray}
\fl \ddot x^i =
-c^2\,\Gamma^i_{\ 00}
-2\,c\,\Gamma^i_{\ 0j}\,\dot x^j
-\Gamma^i_{jk}\,\dot x^j\,\dot x^k
+\dot x^i\,
\left(
c\,\Gamma^0_{\ 00}
+2\,\Gamma^0_{\ 0j}\,\dot x^j
+{1\over c}\,\Gamma^0_{jk}\,\dot x^j\,\dot x^k
\right)\,.
\label{exact_20}
\end{eqnarray}
\noindent
Substituting the Christoffel symbols one gets the differential
equations for the light propagation in metric (\ref{exact_5}):
\begin{eqnarray}
\fl \ddot{\ve{x}} &=&
\frac{a}{x^2} \left[ - c^2 \frac{1 - a}{(1 + a)^3} - \dot{\ve{x}} \cdot \dot{\ve{x}}
+ a \frac{2 - a}{1 - a^2} \left( \frac{ {\ve{x}} \cdot \dot{\ve{x}}}{x} \right)^2
\right] \ve{x}
 +  2 \frac{a}{x^2} \;\frac{2 - a}{1 - a^2}
( \ve{x} \cdot \dot{\ve{x}}) \, \dot{\ve{x}} \,.
\label{exact_25}
\end{eqnarray}
\noindent
Eq. (\ref{isotropic_15}) for the isotropic condition together with
$\dot{\ve{x}}\cdot\dot{\ve{x}} = c^2\,s^2$ could be used to avoid the
term containing $\dot{\ve{x}} \cdot \dot{\ve{x}}$, but this does not
simplify the equations and we prefer not to do this here.

\section{Numerical integration of the equations of light propagation}
\label{Section:numerical_integration}

Our goal is to integrate (\ref{exact_25}) numerically to get a
solution for the trajectory of a light ray with an accuracy much
higher than the goal accuracy of $1\ \muas\approx 4.8\times 10^{-12}$ radians.
For these numerical integrations a simple FORTRAN 95 code using
quadruple (128 bit) arithmetic has been written. Numerical integrator
ODEX \cite{ODEX} has been adapted to the quadruple precision. ODEX is
an extrapolation algorithm based on the explicit midpoint rule. It has
automatic order selection, local accuracy control and dense output.
Using forth and back integration to estimate the accuracy, each
numerical integration is automatically checked to achieve a numerical
accuracy of at least $10^{-24}$ in the components of both position and
velocity of the photon at each moment of time.

The numerical integration is first used to solve the initial-value
(Cauchy) problem for differential equations (\ref{exact_25}).
Eq. (\ref{isotropic_15}) should be used to choose the initial
conditions. The problem of light propagation has thus only 5 degrees
of freedom: 3 degrees of freedom correspond to the position of the
photon and two other degrees of freedom correspond to the unit
direction of light propagation (of course, in the Schwarzschild field
with its symmetry one has also further integrals of motion, 
but here we ignore this; see Section
\ref{section-impact-parameters} below). The absolute value of the coordinate light
velocity can be computed from (\ref{isotropic_15}). Fixing initial
position of the photon $\ve{x}(t_0)$ and initial (unit) direction of
propagation $\ve{\mu}$ one gets the initial velocity of the photon as
function of $\ve{\mu}$ and $s$ computed for given $\ve{\mu}$ and
$\ve{x}$ as given by (\ref{isotropic_15}):
\begin{eqnarray}
\ve{x}(t_0) &=& \ve{x}_0 \,,
\nonumber\\
\dot{\ve{x}} (t_0) &=&  c\, s \,\ve{\mu} \,.
\label{num_5}
\end{eqnarray}
\noindent
The numerical integration yields the position $\ve{x}$ and velocity
$\dot{\ve{x}}$ of the photon as function of time $t$. The dense output
of ODEX allows one to obtain the position and velocity of the photon
on a selected grid of moments of time.  Eq.~(\ref{isotropic_15}) must
hold for any moment of time as soon as it is satisfied by the initial
conditions. Therefore, Eq.~(\ref{isotropic_15}) can also be used to check
the accuracy of numerical integration.

For the purposes of this work we need to have an accurate solution of
two-value boundary problem. That is, a solution of
Eq. (\ref{exact_25}) with boundary conditions
\begin{eqnarray}
\ve{x} (t_\source) &=& \ve{x}_\source ,
\nonumber\\
\ve{x} (t_\obs) &=& \ve{x}_\obs \, ,
\label{num_10}
\label{boundary-problem}
\end{eqnarray}
\noindent
where $\ve{x}_\source$ and $\ve{x}_\obs$ are two given constants,
$t_\source$ is assumed to be fixed and $t_\obs$ is unknown and should
be determined by solving (\ref{exact_25}). Instead of using some
numerical methods to solve this boundary problem directly, we generate
solutions of a family of boundary problems from our solution of the
initial value problem (\ref{num_5}). Each intermediate result computed
during the numerical integration with initial conditions (\ref{num_5})
gives us a high-accuracy solution of the corresponding two-value
boundary problem (\ref{num_10}): $t_\obs$ and $\ve{x}_\obs$ are simply
taken from the numerical integration.

\begin{figure}
\caption{\label{fig:definitions}
Definitions of vectors $\ve{x}_\obs$, $\ve{x}_\source$,
$\ve{k}$, $\ve{n}$, $\ve{\sigma}$. Vectors $\ve{d}$ (defined in Section
\ref{section-equations-pN}) 
and $\ve{d}_\sigma$ (defined in Section \ref{section-impact-parameters}) 
are also shown.   
}
\vspace{10pt}
\begin{indented}
\item[]
\includegraphics[scale=0.78]{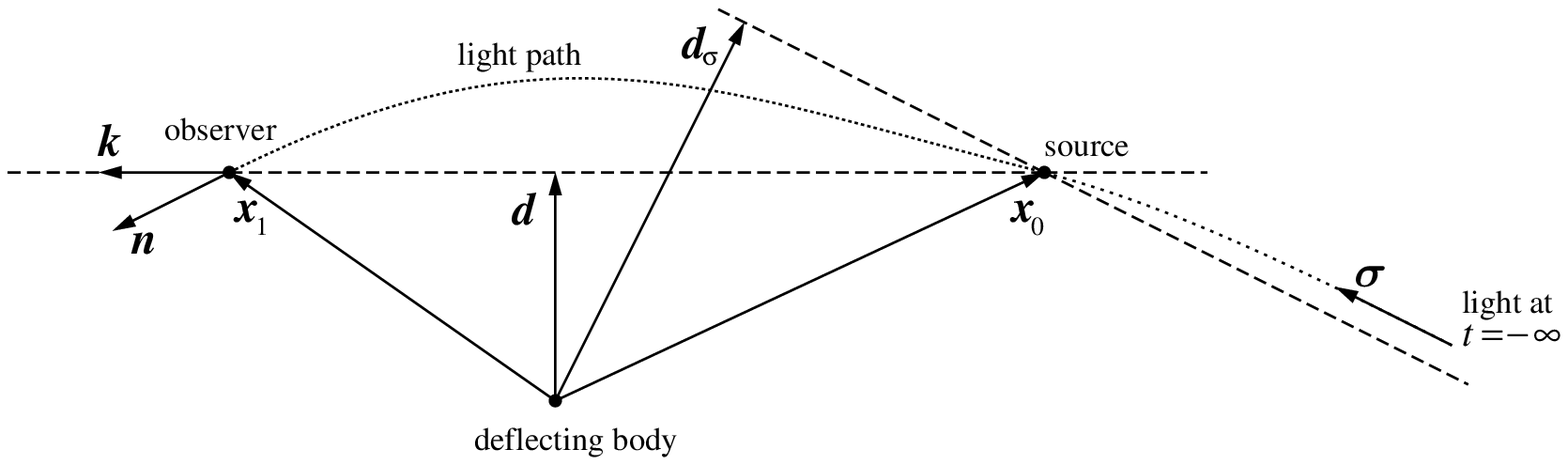}
\end{indented}
\end{figure}

As discussed in \cite{Klioner2003}, the light propagation is
characterized by three unit vectors (see Figure \ref{fig:definitions}):
the coordinate direction $\ve{n}$ of light propagation at the point of
reception
\begin{equation}
  \label{eq:n-numerical}
  \ve{n}={\dot{\ve{x}}(t_\obs)\over\left|\dot{\ve{x}}(t_\obs)\right|}\,,
\end{equation}
\noindent
the coordinate direction $\ve{\sigma}$ of light propagation for time going
to minus infinity, 
\begin{equation}
  \label{eq:sigma-definition}
  \ve{\sigma}=\lim_{t\to-\infty}{1\over c}\,{\dot{\ve{x}}}(t)\,,
\end{equation}
\noindent
and the coordinate direction $\ve{k}$ from the
point of light emission to the point of reception
\begin{equation}
  \label{eq:k-definition}
  \ve{k}={\ve{R}\over R},\qquad \ve{R}=\ve{x}_\obs-\ve{x}_\source.
\end{equation}

In the following discussion we will compare predictions of various
analytical models for $\ve{n}$ in the framework of the boundary
problem (\ref{num_10}). The reference value for these comparisons can
be computed using (\ref{eq:n-numerical}) and $\dot{\ve{x}}(t_\obs)$
from the numerical integration.  The accuracy of this $\ve{n}$
computed from our numerical integrations is guaranteed to be of the
order of $10^{-24}$ radians and can be considered as exact for our
purposes.

\section{The deficiency of the standard post-Newtonian approach}
\label{section-standard-pN}

Let us now demonstrate that the standard post-Newtonian formulas for
the light propagation have too large numerical errors when compared to
the accurate numerical solution of the geodetic equations described in
the previous Section.

\subsection{Equations of the post-Newtonian approach}
\label{section-equations-pN}

The well-known equations of light propagation in first post-Newtonian
approximation with PPN parameters have been discussed by many authors
(see, for example, \cite{Will1993,Brumberg1991,Soffel1989}).
Let us here summarize the standard post-Newtonian formulas.
The differential equations for the light rays read
(see also Section \ref{section-equations} below)
\begin{eqnarray}
\ddot{\ve{x}} &=&
- \, \left(c^2 + \gamma \, \dot x^k\,\dot x^k\right)\,{a\,\ve{x}\over x^2}
+ 2 \, (1 + \gamma ) \,
{a\,\dot{\ve{x}}\,(\dot x^k\,x^k)\over x^2} + {\cal O}(c^{-2}) \,.
\label{pN_15}
\end{eqnarray}
\noindent
The analytical solution of (\ref{pN_15}) can be written in the form
\begin{eqnarray}
\ve{x}(t) &=& \ve{x}_{\rm pN} + {\cal O} (c^{- 4}) \,,
\label{pN_20}
\\
\ve{x}_{\rm pN} &=& \ve{x}_0 + c \, (t - t_0) \, \ve{\sigma} + \Delta\ve{x}(t) \,,
\label{pN_21}
\end{eqnarray}
\noindent
where
\begin{eqnarray}
\fl \Delta \ve{x} (t) = - (1 + \gamma) m  \left( \ve{\sigma} \times
(\ve{x}_0 \times \ve{\sigma}) \left( \frac{1}{x - \ve{\sigma} \cdot \ve{x}}
- \frac{1}{x_0 - \ve{\sigma} \cdot \ve{x}_0} \right) + \ve{\sigma}\,
\log \frac{x + \ve{\sigma} \cdot \ve{x}}
{x_0 + \ve{\sigma} \cdot \ve{x}_0} \right).
\nonumber\\
\label{pN_25}
\end{eqnarray}
\noindent
Solution (\ref{pN_20})--(\ref{pN_25}) satisfies the following 
initial conditions:
\begin{eqnarray}
&&\ve{x}(t_0) = \ve{x}_0\,,
\nonumber\\
&&\lim\limits_{t \rightarrow -\infty}\,\dot{\ve{x}}(t) = c\,\ve{\sigma} \,.
\label{cauchy_5}
\end{eqnarray}
\noindent
From (\ref{pN_20})--(\ref{pN_25}) it is easy to derive the following
expression for the unit tangent vector at the observer's position
$\ve{x}_\obs$ for the boundary problem (\ref{boundary-problem}) (the
standard technique to do this is given, e.g.  in \cite{Brumberg1991}
and used below in Section \ref{section-boundary-problem} in the
post-post-Newtonian approximation):
\begin{eqnarray}
\ve{n}_{\rm pN}
&=& \ve{k} - (1 + \gamma)\,m\,{\ve{d}\over d^2}\,
{x_\source x_\obs-\ve{x}_\source\cdot\ve{x}_\obs\over x_\obs R}\,\,,
\label{pN_30}
\end{eqnarray}
\noindent
where $\ve{R}$ and $\ve{k}$ are defined by (\ref{eq:k-definition}), and
$\ve{d} = \ve{k} \times (\ve{x}_{\source} \times \ve{k}) =
\ve{k} \times (\ve{x}_{\obs} \times \ve{k})$ is the impact parameter
of the straight line connecting $\ve{x}_\source$ and $\ve{x}_\obs$.

\subsection{Comparison of the post-Newtonian 
formula and the numerical solution}
\label{Sec:numerical-comparison}

In order to investigate the accuracy of the standard post-Newtonian
formulas we have compared the post-Newtonian predictions of the light
deflection with the results of the numerical solution of geodetic
equations. Here, we calculate the angle between the unit tangent
vector $\ve{n}_{\rm pN}$ defined by (\ref{pN_30}) and the vector
$\ve{n}$ computed using (\ref{eq:n-numerical}) from the numerical
integration of (\ref{exact_25}).

\begin{figure}
\caption{\label{fig:numeric1}
The angle between 
$\ve{n}_{\rm pN}$ and $\ve{n}$ for Jupiter.
The vector $\ve{n}_{\rm pN}$ is evaluated by means of the
standard post-Newtonian formula (\ref{pN_30}), while
$\ve{n}$ is taken from the numerical integration as
described in Section \ref{Section:numerical_integration}. 
Impact parameter $d$ is taken to be the radius of Jupiter and
the distance $x_\obs$ between Jupiter and the observer is 6 au.}
\begin{indented}
\item[]
\includegraphics[scale=0.3,angle=-90]{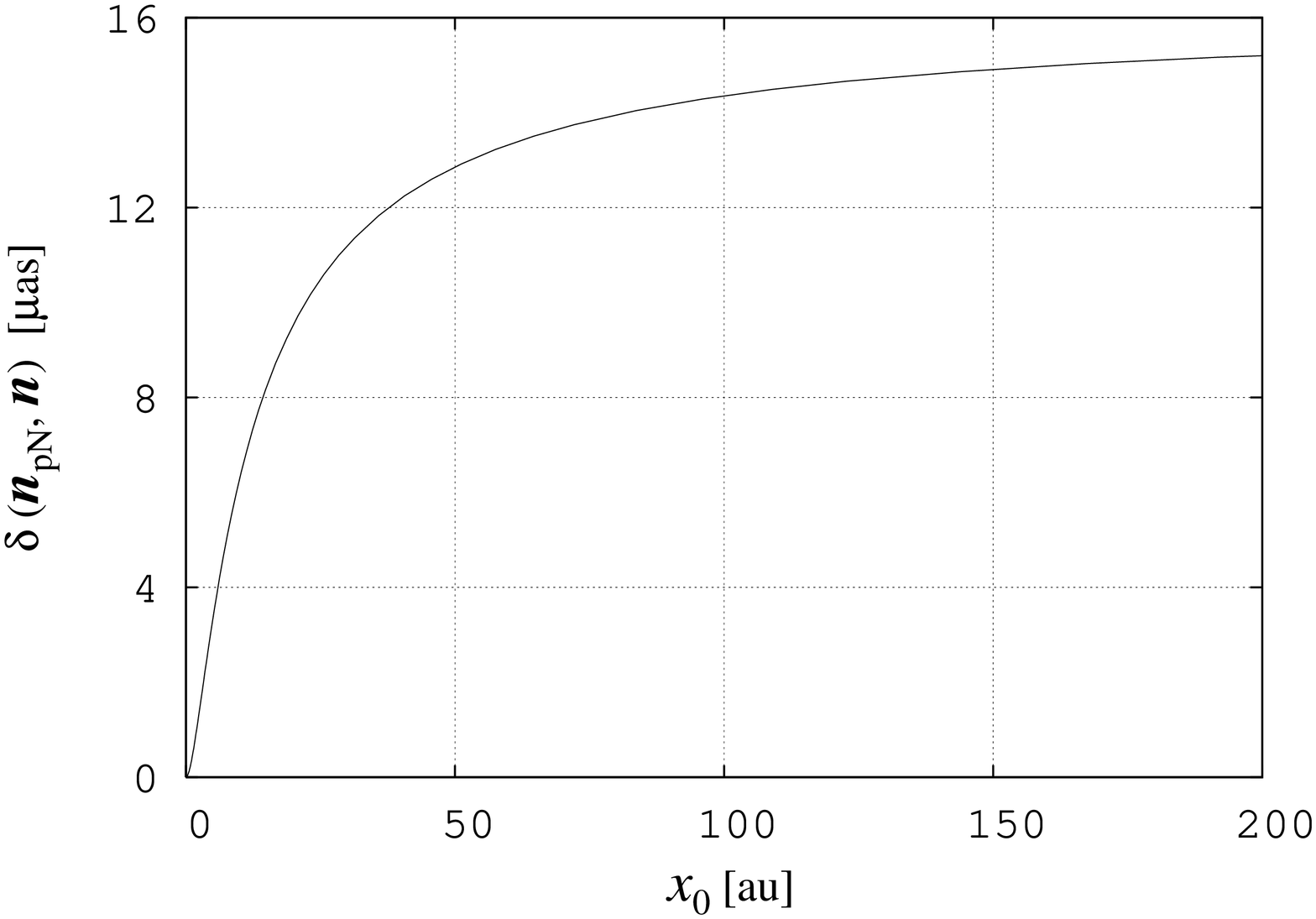}
\end{indented}
\end{figure}

Having performed extensive tests, we have found that, in the real
solar system, the error of $\ve{n}_{\rm pN}$ for observations made by
an observer situated in the vicinity of the Earth attains 16
\muas. These results are illustrated by Table~\ref{table0} and
Figure~\ref{fig:numeric1}. Table~\ref{table0} contains the parameters we
have used in our numerical simulations as well as the maximal angular
deviation between $\ve{n}_{\rm pN}$ and $\ve{n}$ in each set of
simulations. We have performed simulations with different bodies of
the solar systems, assuming that the minimal impact distance $d$ is
equal to the radius of the corresponding body, and the maximal
distance $x_\obs=|\ve{x}_\obs|$ between the gravitating body and the
observer is given by the maximal distance between the gravitating 
body and the Earth.  The simulation shows that the error of
$\ve{n}_{\rm pN}$ is generally increasing for larger $x_\obs$ and
decreasing for larger $d$.  The dependence of the error of
$\ve{n}_{\rm pN}$ for fixed $d$ and $x_\obs$ and increasing distance $x_\source$
between the gravitating body and the source is given on
Figure~\ref{fig:numeric1} for the case of Jupiter, where minimal $d$ and
maximal $x_\obs$ (according to Table~\ref{table0}) were used. Moreover, the
error of $\ve{n}_{\rm pN}$ is found to be proportional to $m^2$ which
leads us to the necessity to deal with the post-post-Newtonian
approximation for the light propagation.

\begin{table}
  \caption{\label{table0}
    Numerical parameters of the Sun and giant planets are taken from
    \cite{Encyclopedia, IERS2003}. 
    $d_{\rm min}$ is the minimal value of the impact parameter $d$ that was used
    in the simulations.
    For each body, $d_{\rm min}$ is equal
    the radius of the body. For the Sun at $45^\circ$ the impact parameter is computed as
    $d = \sin45^\circ \times 1 \, {\rm au}$.
    $x_\obs^{\rm max}$ is the maximal absolute value of the distance $x_\obs$ between 
    the gravitating body and the observer 
    that was used in the simulations.
    $\delta_{\rm max}$ is the maximal angle between $\ve{n}_{\rm pN}$ and $\ve{n}$ 
    found in our numerical simulations.}
\begin{indented}
\item[]
\begin{tabular}{@{}rllllll}
\br
&{Sun}
&{Sun at $45^{\circ}$}
&{Jupiter}
&{Saturn}
&{Uranus}
&{Neptune}\\
\mr
&&&&&&\\[-10pt]
$m=G M/c^2$\ [m] & 1476.6 & 1476.6  & 1.40987  & 0.42215 & 0.064473 & 0.076067 \\
$d_{\rm min}$\ [$10^6$ m] & 696.0 & 105781.7 & 71.492 & 60.268 & 25.559 & 24.764 \\
$x_1^{\rm max}$\ [au] & 1  & 1 & 6 & 11 & 21 & 31 \\
\mr
&&&&&&\\[-10pt]
$\delta_{\rm max}$\ [\muas]&3187.8 & $6.32 \times 10^{-4}$ & 16.13 & 4.42 & 2.58 & 5.84 \\
\br
\end{tabular}
\end{indented}
\end{table}

\section{Analytical post-post-Newtonian solution}
\label{section-ppN-solution}

The goal of this Section is to derive a rigorous analytical
post-post-Newtonian solution for light propagation in the
gravitational field of one spherically symmetric body in the framework
of the PPN formalism extended by a non-linear parameter for the
terms of order $c^{-4}$ in $g_{ij}$. The geodetic equation for the
light ray in Schwarzschild metric can in principle be integrated
exactly \cite{Chandrasekhar1983}. However, such an analytical
solution is given in terms of elliptic integrals and is not very
suitable for massive calculations. Besides that, only the trajectory
of the photon is readily available from the literature, but not the
position and velocity of a photon as functions of time.  Fortunately,
in many cases of interest approximate solutions are sufficient. The
standard way to solve the geodetic equation is the well-known
post-Newtonian approximation scheme. Normally, in practical
applications of relativistic light propagation, the first
post-Newtonian solution is used. Post-post-Newtonian effects have been
also sometimes considered \cite{Hellings1986,Moyer:2000}, but in a way
which cannot be called self-consistent since no rigorous solution in
the post-post-Newtonian approximation has been used. Such a rigorous
post-post-Newtonian analytical solution for light propagation in the
Schwarzschild metric has been derived in
\cite{Brumberg1987,Brumberg1991} in general relativity in a class of
gauges. However, the parametrization in
\cite{Brumberg1987,Brumberg1991} does not allow one to consider
alternative theories of gravity and therefore, a post-post-Newtonian
solution for light propagation within the PPN formalism and its
extension to the second post-Newtonian approximation is not
known. However, it is clearly necessary to have such a solution. 
Therefore, our goal is to generalize the
post-post-Newtonian solution of \cite{Brumberg1987} and to extend it
for the boundary problem for light propagation.

\subsection{Differential equations of light propagation and their integral}

The first part of the problem is to derive the differential equations
of light propagation with PPN and post-linear parameters.

\subsubsection{Metric tensor in the parametrized post-post-Newtonian approximation.}

Expanding metric (\ref{exact_5})
in powers of $c^{-1}$, retaining only the terms relevant for the
post-post-Newtonian solution for the light propagation, and introducing
the PPN parameters $\beta$ and $\gamma$ \cite{Will1993} and the
post-linear parameter $\epsilon$ one gets
\begin{eqnarray}\label{g00}
g_{00}&=&-1+2\,a-2\,\beta\,a^2+\OO6,
\nonumber
\\
\label{g0i}
g_{0i}&=&0,
\nonumber
\\
\label{gij}
g_{ij}&=&\delta_{ij}+2\,\gamma\,a\,\delta_{ij}
+\epsilon\,\left(\delta_{ij}+{x^i\,x^j\over x^2}\right)\,a^2+\OO6 \,,
\label{ppN_covariant}
\end{eqnarray}
\noindent
$a$ being again defined by (\ref{a-def}).  
In general relativity one has $\beta=\gamma=\epsilon=1$.
Parameter $\epsilon$ should be considered as a formal way to trace, in
the following calculations, the terms coming from the terms $c^{-4}$
in $g_{ij}$. No physical meaning of $\epsilon$ is claimed here.
However, this parameter is equivalent to parameter $\Lambda$ of
\cite{richtermatzner1,richtermatzner2,richtermatzner3} and parameter
$\epsilon$ of \cite{EpsteinShapiro1980}.

The corresponding contravariant components of metric tensor can be
deduced from (\ref{ppN_covariant}) and are given by
\begin{eqnarray}
g^{00}&=&-1-2\,a+2\,(\beta-2)\,a^2+\OO6,
\nonumber
\\
g^{0i}&=&0,
\nonumber
\\
g^{ij}&=&\delta_{ij}-2\,\gamma\,a\,\delta_{i j}
+\left((4\gamma^2-\epsilon)\,\delta_{ij}
-\epsilon\,{x^i\,x^j\over x^2}\right)\,a^2 + \OO6.
\label{ppN_contravariant}
\end{eqnarray}
\noindent
The determinant of metric tensor reads
\begin{eqnarray}
\label{detg}
g&=&-1-2\,(3\,\gamma-1)\,a
-2\,(\beta+2\,\epsilon+6\,\gamma\,(\gamma-1))\,a^2+\OO6,
\\
\label{sqrt-detg}
\sqrt{-g}&=&1+(3\,\gamma-1)\,a
+(2\,\beta+4\,\epsilon-1+3\,\gamma\,(\gamma - 2))\,a^2 + \OO6.
\end{eqnarray}
\noindent
Metric (\ref{gij}) is obviously harmonic for $\gamma=\beta=\epsilon=1$
since the harmonic conditions (\ref{harmonic-conditions}) take the form
\begin{eqnarray}
{\partial\left(\sqrt{-g }\,g^{0\alpha}\right)\over \partial x^\alpha}&=&0,
\nonumber
\\
{\partial \left(\sqrt{-g}\,g^{i\alpha}\right)\over \partial x^\alpha}&=&
(1-\gamma)\,{a\,x^i\over x^2}
+((1+\gamma)^2-2\beta-2\epsilon)\,{a^2 \, x^i \over x^2}
+\OO6.
\end{eqnarray}

\subsubsection{Christoffel symbols.}

The Christoffel symbols of second kind defined by (\ref{christoffel})
can be derived from metric (\ref{ppN_covariant})--(\ref{ppN_contravariant}):
\begin{eqnarray}
\label{G000}
\fl \Gamma^0_{\ 00} = 0 ,
\\
\label{G00i}
\fl \Gamma^0_{\ 0i} = {a\,x^i\over x^2} + (1-\beta)\,{2\,a^2\,x^i\over x^2}+ \OO6 ,
\\
\label{G0ik}
\fl \Gamma^0_{\ ik} = 0 ,
\\
\label{Gi00}
\fl \Gamma^i_{\ 00} = {a\,x^i\over x^2}
- (\beta + \gamma) \, {2\, a^2\, x^i \over x^2}+\OO6 ,
\\
\label{Gi0k}
\fl \Gamma^i_{\ 0k} = 0 ,
\\
\label{Gikl}
\fl \Gamma^i_{\ kl} = \gamma\,\left(x^i\,\delta_{kl}-x^k\,\delta_{il}-x^l\,
\delta_{ik}\right)\,{a \over x^2}
\nonumber\\
\fl 
\phantom{\Gamma^i_{\ kl} =}
+ \left(2\,(\epsilon-\gamma^2)\,x^i\,\delta_{kl}-(\epsilon-2\,\gamma^2)\,
\left(x^k\,\delta_{il}+x^l\,\delta_{ik}\right)
-2\,\epsilon\,{x^i\,x^k\,x^l\over x^2}
\right)\, {a^2 \over x^2}
+\OO6 .
\end{eqnarray}

\subsubsection{Isotropic condition for the null geodetic.}

From now on, $x^{\alpha}$ denote the coordinates of a photon, $x^i$
denote the spatial coordinates of the photon, and $x=|\ve{x}|$ is the
distance of the photon from the gravitating body that is situated at
the origin of the used reference system. 
As it was discussed in Section \ref{section-isotropic-condition},
Eq. (\ref{isotropic_10}) 
allows one to compute the absolute value of coordinate
velocity of light if the position of the
photon $x^i$ and the unit coordinate direction of its propagation $\mu^i$
($\ve{\mu}\cdot\ve{\mu}=1$)
are given. Using (\ref{ppN_covariant}) 
for $s=|\dot{\ve{x}}|/c$ one gets
\begin{equation}
\label{s-ppN}
\fl s=1-(1+\gamma) \, a + {1\over 2}\,\left(-1+2\,\beta-\epsilon+\gamma\,
(2+3\gamma)-\epsilon\,{\left({\ve{\mu}\cdot\ve{x}\over x}\right)}^2\right) \, a^2
+ {\cal O} (c^{-6}) \, .
\label{isotropic_25}
\end{equation}

\subsubsection{Differential equations of light propagation.}
\label{section-equations}

Inserting the Christoffel symbols (\ref{G000})--(\ref{Gikl}) into (\ref{exact_20}), 
one gets the following
equations of light propagation in post-post-Newtonian approximation
\begin{eqnarray}
\label{geodetic-ppN}
\ddot{\ve{x}} &=&
- \left(c^2+\gamma\,\dot{\ve{x}}\cdot\dot{\ve{x}}\right)\,{a\,\ve{x}\over x^2}
+2\,(1+\gamma)\,{a\,\dot{\ve{x}}\,(\dot{\ve{x}}\cdot\ve{x})\over x^2}
\nonumber
\\
&&
+2\,\left((\beta+\gamma)\,c^2+(\gamma^2-\epsilon)\,
(\dot{\ve{x}}\cdot\dot{\ve{x}})\right)\,{a^2\,\ve{x}\over x^2}
+2\,\epsilon\,{a^2\,\ve{x}\,(\dot{\ve{x}}\cdot\ve{x})^2\over x^4}
\nonumber
\\
&&
+2\,(2(1-\beta)+\epsilon-2\,\gamma^2)\,
{a^2\,\dot{\ve{x}}\,(\dot{\ve{x}}\cdot\ve{x})\over x^2} + \OO4 \,.
\end{eqnarray}
\noindent
Here, for estimating the analytical order of smallness of the terms
we take into account that $|\dot{\ve{x}}| = {\cal O}(c)$. Using
(\ref{s-ppN}) and $\dot{\ve{x}}\cdot\dot{\ve{x}}=c^2\,s^2$
one can simplify (\ref{geodetic-ppN}) to get
\begin{eqnarray}
\label{geodetic-ppN-final}
\ddot{\ve{x}} &=&
-(1+\gamma)\,c^2\,{a\,\ve{x}\over x^2}
+2\,(1+\gamma)\,{a\,\dot{\ve{x}}\,(\dot{\ve{x}}\cdot\ve{x})\over x^2}
\nonumber
\\
&&
+2\,c^2\,\left(\beta-\epsilon+2\,\gamma\,(1+\gamma)\right)\,
{a^2\,\ve{x}\over x^2} + 2\,\epsilon\,{a^2\,\ve{x}\,(\dot{\ve{x}}\cdot\ve{x})^2\over x^4}
\nonumber
\\
&&
+2\,(2(1-\beta)+\epsilon-2\,\gamma^2)\,
{a^2\,\dot{\ve{x}}\,(\dot{\ve{x}}\cdot\ve{x})\over x^2} + \OO4 \,.
\end{eqnarray}

\subsubsection{Equations of light propagation with additional trace
parameter $\alpha$.}
\label{section-equations-with-alpha}

For our purposes it is advantageous to have one more 
parameter that can be used to trace terms in the following calculations
which come from the post-post-Newtonian terms in the equations of
motion of a photon. We denote this parameter $\alpha$ and introduce
it in the above equation simply as a factor for all the
post-post-Newtonian terms in the right-hand side:
\begin{eqnarray}
\ddot{\ve{x}} &=&
-(1+\gamma)\,c^2\,{a\,\ve{x}\over x^2}
+2\,(1+\gamma)\,{a\,\dot{\ve{x}}\,(\dot{\ve{x}}\cdot\ve{x})\over x^2}
\nonumber
\\
&&
+2\,c^2\,\alpha\,\left(\beta-\epsilon+2\,\gamma\,
(1+\gamma)\right)\,{a^2\,\ve{x}\over x^2}
+2\,\alpha\,\epsilon\,{a^2\,\ve{x}\,(\dot{\ve{x}}\cdot\ve{x})^2\over x^4}
\nonumber
\\
&&
+2\,\alpha\,(2(1-\beta)+\epsilon-2\,\gamma^2)\,
{a^2\,\dot{\ve{x}}\,(\dot{\ve{x}}\cdot\ve{x})\over x^2}
+\OO4\,.
\label{geodetic-ppN-final-with-alpha}
\end{eqnarray}
\noindent
Setting $\alpha=0$ in the solution of 
(\ref{geodetic-ppN-final-with-alpha}) one can formally get a
second-order solution for the post-Newtonian equations of light
propagation. The merit of this parameter will be clear below.

\subsection{Initial value problem}
\label{section-initial-value-problem}

Let us now solve analytically an initial value problem for the derived
equations.  For initial conditions (\ref{cauchy_5}) using the same
approach as in \cite{Brumberg1987,Brumberg1991}, one gets:
\begin{eqnarray}
\label{rN-rpN-rppN}
{1\over c}\,\dot{\ve{x}}_N&=&\ve{\sigma},
\label{brumberg_5}
\\
\ve{x}_N&=&\ve{x}_0+c\,(t-t_0)\,\ve{\sigma},
\label{brumberg_10}
\\
{1\over c}\,\dot{\ve{x}}_{\rm pN}&=&\ve{\sigma}+m\,\ve{A}_1(\ve{x}_N),
\label{brumberg_15}
\\
\ve{x}_{\rm pN}&=&\ve{x}_N+m\,\left(\ve{B}_1(\ve{x}_N)-\ve{B}_1(\ve{x}_0)\right),
\label{brumberg_20}
\\
{1\over c}\,\dot{\ve{x}}_{\rm ppN}&=&\ve{\sigma}+m\,\ve{A}_1(\ve{x}_{\rm pN})
+m^2\,\ve{A}_2(\ve{x}_N),
\label{brumberg_25}
\\
\ve{x}_{\rm ppN}&=&\ve{x}_N+m\,\left(\ve{B}_1(\ve{x}_{\rm pN})-\ve{B}_1(\ve{x}_0)\right)
+m^2\,\left(\ve{B}_2(\ve{x}_N)-\ve{B}_2(\ve{x}_0)\right),
\label{brumberg_30}
\end{eqnarray}
\noindent
where
\begin{eqnarray}
\fl \ve{A}_1(\ve{x}) = - (1+\gamma)\,
\left({{\ve{\sigma}\times(\ve{x}\times\ve{\sigma})\over x
(x -\ve{\sigma}\cdot\ve{x})}+{\ve{\sigma}\over x}}\right),
\label{brumberg_35}
\\
\fl \ve{B}_1(\ve{x}) = -(1+\gamma)\,
\left({\ve{\sigma}\times(\ve{x}\times\ve{\sigma})\over x
-\ve{\sigma}\cdot\ve{x}}
+\ve{\sigma}\,\log {( x +\ve{\sigma}\cdot\ve{x})}\right),
\label{brumberg_40}
\\
\label{A_2a}
\fl \ve{A}_2(\ve{x}) =
-{1\over 2}\,\alpha\,\epsilon\,{\ve{\sigma}\cdot\ve{x}\over x^4}\,\ve{x}
+2\,(1+\gamma)^2\,{\ve{\sigma}\times(\ve{x}\times\ve{\sigma})\over x^2\,
\left( x - \ve{\sigma}\cdot\ve{x}\right)}
+(1+\gamma)^2\,{\ve{\sigma}\times(\ve{x}\times\ve{\sigma})\over x \,
{\left(x -\ve{\sigma}\cdot\ve{x}\right)}^2}
\nonumber
\\
\fl - (1+\gamma)^2\,{\ve{\sigma}\over x\,\left(x-\ve{\sigma}\cdot\ve{x}\right)}
+\left(2(1-\alpha+\gamma)\,(1+\gamma)+\alpha\,
\beta-{1\over 2}\,\alpha\,\epsilon\right)\,{\ve{\sigma}\over x^2}
\nonumber
\\
\fl -{1\over 4}\,\left(8\,(1+\gamma-\alpha\,\gamma)\,(1+\gamma)-4\,\alpha\,\beta+3\,\alpha\,\epsilon\right)\,
\left(\ve{\sigma}\cdot\ve{x}\right)\,
{\ve{\sigma}\times(\ve{x}\times\ve{\sigma})\over x^2\,|\ve{\sigma}\times\ve{x}|^2}
\nonumber
\\
\fl -{1\over 4}\,\left(8\,(1+\gamma-\alpha\,\gamma)\,(1+\gamma)-4\,\alpha\,\beta+3\,\alpha\,\epsilon\right)\,
{\ve{\sigma}\times(\ve{x}\times\ve{\sigma})\over |\ve{\sigma}\times\ve{x}|^3}\,
\left( \pi - \delta (\ve{\sigma}, \ve{x}) \right) \,,
\label{brumberg_45}
\\
\fl \ve{B}_2(\ve{x}) = 
-(1+\gamma)^2\,{\ve{\sigma}\over x - \ve{\sigma}\cdot\ve{x}}
+(1+\gamma)^2\,{\ve{\sigma}\times(\ve{x}\times\ve{\sigma})\over {\left( x - \ve{\sigma}\cdot\ve{x}\right)}^2}
+{1\over 4}\,\alpha\,\epsilon\,{\ve{x}\over x^2}
\nonumber
\\
\fl-{1\over 4}\,\alpha\,\left(8\,(1+\gamma)-4\,\beta+3\,\epsilon\right)\,{\ve{\sigma}\over |\ve{\sigma}\times\ve{x}|}\,
\left( \frac{\pi}{2} - \delta (\ve{\sigma}, \ve{x}) \right)
\nonumber
\\
\fl-{1\over 4}\,\left(8\,(1+\gamma-\alpha\,\gamma)\,(1+\gamma)-4\,\alpha\,\beta+3\,\alpha\,\epsilon\right)\,\left(\ve{\sigma}\cdot\ve{x}\right)\,
{\ve{\sigma}\times(\ve{x}\times\ve{\sigma})\over |\ve{\sigma}\times\ve{x}|^3}\,
\left( \pi - \delta (\ve{\sigma}, \ve{x}) \right),
\label{brumberg_50}
\end{eqnarray}
\noindent
or, alternatively, for $\ve{B}_1$ and $\ve{B}_2$
\begin{eqnarray}
\label{B_1a}
\fl \ve{B}_1(\ve{x}) = -(1+\gamma)\,
\left({\ve{\sigma}\times(\ve{x}\times\ve{\sigma})\over x -
\ve{\sigma}\cdot\ve{x}}
-\ve{\sigma}\,\log {(x-\ve{\sigma}\cdot\ve{x})}\right),
\\
\label{B_2}
\fl \ve{B}_2(\ve{x}) =
+(1+\gamma)^2\,{\ve{\sigma}\over x -\ve{\sigma}\cdot\ve{x}}
+(1+\gamma)^2\,{\ve{\sigma}\times(\ve{x}\times\ve{\sigma})\over
{\left(x-\ve{\sigma}\cdot\ve{x}\right)}^2}
+{1\over 4}\,\alpha\,\epsilon\,{\ve{x}\over x^2}
\nonumber
\\
\fl -{1\over 4}\,\alpha\,\left(8\,(1+\gamma)-4\,\beta+3\,\epsilon\right)\,{\ve{\sigma}\over |\ve{\sigma}\times\ve{x}|}\,
\left( \frac{\pi}{2} - \delta (\ve{\sigma}, \ve{x}) \right)
\nonumber
\\
\fl -{1\over 4}\,\left(8\,(1+\gamma-\alpha\,\gamma)\,(1+\gamma)-4\,\alpha\,\beta+3\,\alpha\,\epsilon\right)\,
\left(\ve{\sigma}\cdot\ve{x}\right)\,
{\ve{\sigma}\times(\ve{x}\times\ve{\sigma})\over |\ve{\sigma}\times\ve{x}|^3}\,
\left( \pi - \delta (\ve{\sigma}, \ve{x}) \right).
\end{eqnarray}
\noindent
With these definitions the solution of
(\ref{geodetic-ppN-final-with-alpha}) reads
\begin{eqnarray}
\ve{x} (t) &=& \ve{x}_{\rm ppN} (t) + \OO6,
\nonumber\\
\frac{1}{c} \dot{\ve{x}} (t) &=& 
\frac{1}{c} \dot{\ve{x}}_{\rm ppN} (t) + \OO6.
\end{eqnarray}
It is easy to check that the solution for coordinate velocity of light
$\dot{\ve{x}}_{\rm ppN}$ satisfies the integral (\ref{s-ppN}). In
order to demonstrate this fact, it is important to understand that
position $\ve{x}$ in (\ref{s-ppN}) lies on the trajectory of the
photon and must be therefore considered as $\ve{x}_{\rm pN}$ in the
post-Newtonian terms and as $\ve{x}_{\rm N}$ in the
post-post-Newtonian terms of (\ref{brumberg_25}).

\subsection{Vector $\ve{n}$ in the initial problem}

Using (\ref{brumberg_25}) one gets 
\begin{equation}
\ve{n}=\ve{\sigma} + m\,\ve{C}_1(\ve{x}_{\rm pN}) + m^2\,\ve{C}_2(\ve{x}_{\rm N})
+ \OO6,
\label{sigma2n-formal}
\end{equation}
\noindent
where
\begin{eqnarray}
\fl \ve{C}_1 (\ve{x}) = \ve{A}_1 (\ve{x})
\,-\,\ve{\sigma} \left(\ve{\sigma} \cdot \ve{A}_1 (\ve{x})\right)
\, = \, - (1 + \gamma) \frac{\ve{\sigma} \times (\ve{x} \times \ve{\sigma})}
{x \, (x - \ve{\sigma} \cdot \ve{x}) } \,,
\nonumber\\
\fl \ve{C}_2 (\ve{x}) = \ve{A}_2 (\ve{x})
\,-\, \ve{A}_1 (\ve{x})\,
\left(\ve{\sigma} \cdot \ve{A}_1 (\ve{x}) \right) \,-\, \frac{1}{2}\,
\ve{\sigma}\,\left(\ve{A}_1(\ve{x})\cdot\ve{A}_1(\ve{x})\right)\,-\,
\ve{\sigma}\,\left(\ve{\sigma} \cdot \ve{A}_2 (\ve{x})\right)
\nonumber\\
\fl 
\phantom{\ve{C}_2(\ve{x})=}
+ \frac{3}{2} \,\ve{\sigma}\,
\left(\ve{\sigma} \cdot \ve{A}_1 (\ve{x}) \right)^2
\nonumber\\
\fl 
\phantom{\ve{C}_2(\ve{x})}
= - \frac{1}{2}\,\alpha\,\epsilon\, \frac{\ve{\sigma} \cdot \ve{x}}{x^4}\,
\ve{\sigma} \times (\ve{x} \times \ve{\sigma})
+\,(1 + \gamma)^2\,\frac{\ve{\sigma} \times (\ve{x} \times \ve{\sigma})}
{x^2\,(x - \ve{\sigma} \cdot \ve{x})}
\nonumber\\
\fl 
\phantom{\ve{C}_2(\ve{x})=}
+(1 + \gamma)^2\,
\frac{\ve{\sigma} \times (\ve{x} \times \ve{\sigma})}
{x\,(x - \ve{\sigma} \cdot \ve{x})^2} \,-\,\frac{1}{2}\,
(1 + \gamma)^2\,\frac{\ve{\sigma}}
{x^2}\,\frac{x + \ve{\sigma} \cdot \ve{x}}{x - \ve{\sigma} \cdot \ve{x}}
\nonumber\\
\phantom{\ve{C}_2(\ve{x})=}
\fl -{1\over 4}\,\left(8\,(1+\gamma-\alpha\,\gamma)\,(1+\gamma)-4\,\alpha\,\beta+3\,\alpha\,\epsilon\right)\,
\left(\ve{\sigma}\cdot\ve{x}\right)\,
{\ve{\sigma}\times(\ve{x}\times\ve{\sigma})\over x^2\,|\ve{\sigma}\times\ve{x}|^2}
\nonumber
\\
\fl
\phantom{\ve{C}_2(\ve{x})=}
-{1\over 4}\,\left(8\,(1+\gamma-\alpha\,\gamma)\,(1+\gamma)-4\,\alpha\,\beta+3\,\alpha\,\epsilon\right)\,
{\ve{\sigma}\times(\ve{x}\times\ve{\sigma})\over |\ve{\sigma}\times\ve{x}|^3}\,
\left( \pi - \delta (\ve{\sigma} , \ve{x}) \right)\,.
\label{n_5}
\end{eqnarray}

\subsection{Impact parameters}
\label{section-impact-parameters}

As we have seen in Sections \ref{section-equations-pN} and
\ref{section-initial-value-problem} the usual analytical solutions are
expressed through one of the two following impact parameters:
\begin{eqnarray}
\ve{d}_{\sigma} &=& \ve{\sigma} \times \left( \ve{x}_0 \times \ve{\sigma}\right),
\label{impact_10}
\\
\ve{d} &=& \ve{k} \times \left( \ve{x}_{\source} \times \ve{k}\right) 
= \ve{k} \times \left( \ve{x}_{\obs} \times \ve{k}\right),
\label{impact_5}
\end{eqnarray}
\noindent
where $\ve{x}_0$ is the initial point in both Cauchy and boundary
problems given by (\ref{cauchy_5}) and (\ref{boundary-problem}),
respectively, while $\ve{x}_{\obs}$ is the final position in the
boundary problem.  Both these impact parameters naturally arise in
practical calculations of light propagation when positions of source
and observer are given in some reference system (e.g., in the BCRS
\cite{Klioner2003}). However, these parameters are clearly
coordinate-dependent and have no profound physical meaning. One can
expect that formulas involving these impact parameters contain some
spurious, non-physical terms obscuring the physical meaning of the formulas.
As we will see below it is indeed the case.  Now, we introduce
another impact parameter
\begin{equation}
\ve{d}^{\;\prime} 
= 
\lim_{t \to-\infty} {1\over c}\,\dot{\ve{x}}(t)\times 
\left(\ve{x}(t)\times{1\over c}\,\dot{\ve{x}}(t)\right) 
= \lim_{t \to -\infty} \ve{\sigma} \times \left( \ve{x}(t) \times \ve{\sigma}\right) .
\label{impact_20}
\label{d-prime-xdot}
\end{equation}
\noindent
For a similar impact parameter defined at $t\to+\infty$
\begin{equation}
\ve{d}^{\;\prime\prime} = 
\lim_{t \to +\infty} {1\over c}\,\dot{\ve{x}}(t) \times \left( \ve{x}(t) \times {1\over c}\,\dot{\ve{x}}(t)\right)=
\lim_{t \to+\infty} \ve{\nu} \times \left( \ve{x}(t) \times \ve{\nu}\right),
\label{d-prime-prime}
\end{equation}
\noindent
where $\ve{\nu}=\lim\limits_{t \to+\infty}{1\over c}\,\dot{\ve{x}}(t)$, 
one has $\left|\ve{d}^{\;\prime}\right|
=\left|\ve{d}^{\;\prime\prime}\right|$. It is also clear that the
angle between $\ve{d}^{\;\prime}$ and $\ve{d}^{\;\prime\prime}$ is
equal to the full light deflection (see below).  Since both
$\ve{d}^{\;\prime}$ and $\ve{d}^{\;\prime\prime}$ reside at time-like
infinity and since the metric under study is
asymptotically flat, these parameters can be called
coordinate-independent.

One can show that $d^{\;\prime}=d^{\;\prime\prime}$ coincides with the
impact parameter $D$ introduced, e.g., by Eq. (215) of Section 20 of
\cite{Chandrasekhar1983} in terms of full energy and angular momentum
of the photon (see also \cite{BodennerWill2003} for a useful
discussion). Indeed, in polar coordinates $(x,\varphi)$ the
Chandrasekhar's impact parameter $D=f(x)\,x^2\,\dot{\varphi}$, where
$\lim\limits_{x\to\infty}f(x)=1$.  Clearly,
$x^2\,\dot{\varphi}=|\dot{\ve{x}}(t)\times\ve{x}(t)|$ and it is
obvious that $d^{\;\prime}=d^{\;\prime\prime}=D$. Interestingly, this
discussion allows one to find an exact integral of the equations of
motion for a photon in the Schwarzschild field. The equations of light
propagation (\ref{exact_25}) in the Schwarzschild metric
(\ref{exact_5}) in harmonic coordinates have an integral
\begin{equation}
\ve{D}={(1+a)^3\over 1-a}\,{1\over c}\,\dot{\ve{x}}(t)\times\ve{x}(t)={\rm const},
\end{equation}
\noindent
while for the parametrized post-post-Newtonian equations of motion given
by (\ref{geodetic-ppN-final-with-alpha}) one has
\begin{eqnarray}
\fl \ve{D}=\exp\biggl(2(1+\gamma)\,a+\alpha\,\left(2\,(1-\beta)+\epsilon-2\gamma^2\right)\,a^2
\biggr)\,{1\over c}\,\dot{\ve{x}}(t)\times\ve{x}(t)
\nonumber\\
\fl =\left(1+2(1+\gamma)\,a+\left(2(1+\gamma)^2+\alpha\,\left(2\,(1-\beta)+\epsilon-2\gamma^2\right)\right
)\;a^2\right)\,{1\over c}\,\dot{\ve{x}}(t)\times\ve{x}(t)+\OO{6}
\nonumber\\
\fl ={\rm const}.
\label{D-ppN}
\end{eqnarray}
\noindent
The first line of (\ref{D-ppN}) represents an {\it exact}\/ integral of the
(approximate) equations of motion (\ref{geodetic-ppN-final-with-alpha}).
In both cases the Chandrasekhar's $D$ is the absolute value of
$\ve{D}$ as given above.

Let us stress that the impact parameter $\ve{d}^{\;\prime}$
is not convenient for practical calculations, but
we will use it below 
to understand the physical origin of various terms in the formulas
describing the light propagation.
Therefore,
we need to have a relation between impact parameters (\ref{impact_10}), (\ref{impact_5}), 
and (\ref{impact_20}). Relation
between $\ve{d}^{\;\prime}$ and $\ve{d}_\sigma$ can be derived 
using the post-Newtonian solution for light propagation 
given above:
\begin{equation}
\ve{d}^{\;\prime} = \ve{d}_{\sigma} \left( 1 + (1+\gamma) \;\frac{m}{d_{\sigma}^2}\;
\left(x_{\source} + \ve{\sigma} \cdot \ve{x}_{\source} \right) \right) 
+ \OO4.
\label{impact_25}
\end{equation}
\noindent
Relation of $\ve{d}^{\;\prime}$ and $\ve{d}$ can be derived using
formulas of Section \ref{section-equations-pN}:
\begin{eqnarray}
\fl
&&
\ve{d}^{\;\prime} =  \ve{d} \left( 1 + \left(1 + \gamma \right) 
\;\frac{m}{d^2}\;
\frac{x_{\obs} + x_{\source}}{R}\;\frac{R^2 - \left(x_{\obs} - x_{\source} \right)^2}{2\;R} \right)
\nonumber\\
&&
\phantom{\ve{d}^{\;\prime} =}
- (1 + \gamma)\;m\;\ve{k}\;\frac{x_{\obs}-x_{\source}+R}{R} 
+ \OO4.
\label{impact_30}
\end{eqnarray}
\noindent
Now we are ready to proceed to the analysis of the post-post-Newtonian
equations of light propagation.

\subsection{Total light deflection}

In order to derive the total light deflection, we have to consider the
limits of the coordinate light velocity $\dot{\ve{x}}$ for $t \rightarrow
\pm \infty$. Using formulas of Section \ref{section-initial-value-problem}
one gets
\begin{eqnarray}
\lim_{t\to-\infty} {1\over c}\,\dot{\ve{x}}(t) &=& \ve{\sigma},
\\
\lim_{t\to+\infty} {1\over c} \, \dot{\ve{x}} (t) &\equiv& \ve{\nu}
\nonumber\\
&& 
\hspace{-1.0cm} 
= \ve{\sigma}  -  2\,(1 + \gamma)\,m\,
\frac{\ve{\sigma} \times (\ve{x}_{\source} \times \ve{\sigma})}{|\ve{x}_{\source} \times \ve{\sigma}|^2}
 -  2 \, (1 + \gamma)^2\,  m^2\,  \frac{\ve{\sigma}}
{|\ve{x}_{\source} \times \ve{\sigma}|^2}
\nonumber\\
&& 
\hspace{-1.0cm} 
-  \frac{1}{4}\, \pi\,
\left(8 (1 + \gamma - \alpha \, \gamma) (1 + \gamma)
- \, 4  \alpha \, \beta + 3 \alpha \, \epsilon \right)\,  m^2\,
\frac{\ve{\sigma} \times (\ve{x}_{\source} \times \ve{\sigma})}{|\ve{x}_{\source} \times \ve{\sigma}|^3}
\nonumber\\
&& 
\hspace{-1.0cm} 
+ 2\, (1 + \gamma)^2\,  m^2  \,\left( x_{\source}  +  \ve{\sigma} \cdot
\ve{x}_{\source} \right)  \frac{\ve{\sigma} \times (\ve{x}_{\source} \times \ve{\sigma})}
{|\ve{x}_{\source} \times \ve{\sigma}|^4} +\OO6.
\label{deflection_5}
\end{eqnarray}
\noindent
Therefore, the total light deflection reads
\begin{eqnarray}
\fl |\ve{\sigma} \times \ve{\nu}| = 2 \, (1 + \gamma) \, m \,
\frac{1}{|\ve{x}_{\source} \times \ve{\sigma}|} \, - \,
2 \, (1 + \gamma)^2 \, m^2 \, ( x_{\source} \, + \ve{\sigma} \cdot \ve{x}_{\source}) \,
\frac{1}{|\ve{x}_{\source} \times \ve{\sigma}|^3}
\nonumber\\
\fl 
\phantom{|\ve{\sigma} \times \ve{\nu}| =}
+ \frac{1}{4}
\left(8 (1 + \gamma - \alpha \, \gamma) (1 + \gamma)
- \, 4 \, \alpha \, \beta + 3 \alpha \, \epsilon \right)\,\pi \, m^2 \,
\frac{1}{|\ve{x}_{\source} \times \ve{\sigma}|^2} +\OO6\,.
\label{deflection_10}
\end{eqnarray}
\noindent
Eq.~(\ref{deflection_10}) defines the sine of the angle of the total light
deflection in post-post-Newtonian approximation. The first term in
(\ref{deflection_10}) is the post-Newtonian expression of total light
deflection. The other two
terms are the post-post-Newtonian corrections. Using $d^\prime$ 
defined by (\ref{d-prime-xdot}) and related to $d_\sigma$ by
(\ref{impact_25})
one can rewrite (\ref{deflection_10}) as
\begin{eqnarray}
\fl |\ve{\sigma} \times \ve{\nu}| = 2 \, (1 + \gamma) \,
\frac{m}{d^\prime}
+ \frac{1}{4}\,
\left(8 (1 + \gamma - \alpha \, \gamma) (1 + \gamma)
- \, 4 \, \alpha \, \beta + 3 \alpha \, \epsilon \right)\,\pi\,
\frac{m^2}{d^{\prime2}} +\OO6\,.
\label{full-deflection-d-prime}
\end{eqnarray}
\noindent
This result with $\alpha=1$ coincides with Eq. (4) of
\cite{EpsteinShapiro1980} and also agrees with the results of
\cite{richtermatzner1,Cowling1984,Brumberg1987,TeyssandierLePoncinLafitte2008}
in the corresponding limits.  It is now clear that the second term in
the right-hand side of (\ref{deflection_10}) ``corrects'' the
main post-Newtonian term converting it to $2(1+\gamma)m/d^\prime$.
Note that the total light deflection $|\ve{\sigma} \times \ve{\nu}|$
is a coordinate-independent quantity and
(\ref{full-deflection-d-prime}) expresses it through
coordinate-independent quantities while (\ref{deflection_10}) does
not.

\section{Post-post-Newtonian solution of the boundary problem}
\label{section-boundary-problem}

For practical modeling of observations it is not sufficient to
consider the initial value problem for light propagation. 
Two-point boundary value problem given by (\ref{boundary-problem})
is important here. This Section is devoted to a derivation 
of the post-post-Newtonian solution of this boundary problem
for (\ref{geodetic-ppN-final-with-alpha}).

\subsection{Formal expressions}

An iterative solution of (\ref{brumberg_5})--(\ref{brumberg_30}) for
the propagation time $\tau=t_\obs-t_\source$ and unit direction $\ve{\sigma}$ reads:
\begin{eqnarray}
c \, \tau  = R &-& m\,\ve{k} \cdot
\left[ \ve{B}_1 (\ve{x}_{\obs}) - \ve{B}_1 (\ve{x}_{\source})\right]
- m^2\,\ve{k} \cdot \left[ \ve{B}_2 (\ve{x}_{\obs}) - \ve{B}_2 (\ve{x}_{\source})\right]
\nonumber\\
&+&\frac{m^2}{2\,R}\,   \left|
\ve{k} \times \left( \ve{B}_1 (\ve{x}_{\obs}) - \ve{B}_1 (\ve{x}_{\source}) \right)
\right|^2+\OO6,
\label{iteration_5}
\end{eqnarray}
\begin{eqnarray}
\ve{\sigma} = \ve{k}&& + m\,\frac{1}{R} \,
\left( \ve{k} \times \left[ \ve{k} \times
(\ve{B}_1 (\ve{x}_{\obs}) - \ve{B}_1 (\ve{x}_{\source})) \right] \right)
\nonumber\\
&&  
+ m^2\,\frac{1}{R} \,
\left( \ve{k} \times \left[ \ve{k} \times
(\ve{B}_2 (\ve{x}_{\obs}) - \ve{B}_2 (\ve{x}_{\source})) \right] \right)
\nonumber\\
&&  + m^2\,\frac{1}{R^2} \,
\left(\ve{B}_1 (\ve{x}_{\obs}) - \ve{B}_1 (\ve{x}_{\source}) \right) \times
 \left[ \ve{k} \times (\ve{B}_1 (\ve{x}_{\obs}) - \ve{B}_1 (\ve{x}_{\source})) \right]
\nonumber\\
&&  - \frac{3}{2}\, m^2\,\frac{1}{R^2} \,\ve{k}\,
\left| \ve{k} \times \left( \ve{B}_1 (\ve{x}_{\obs}) - \ve{B}_1 (\ve{x}_{\source}) \right)
\right|^2+\OO6.
\label{iteration_10}
\end{eqnarray}
\noindent
These expressions are still implicit since in order to achieve the
post-post-Newtonian accuracy one should use the post-Newtonian
relation between $\ve{\sigma}$ and $\ve{k}$ to represent $\ve{\sigma}$
in $\ve{B}_1$ appearing in the post-Newtonian terms. That relation can
be again obtained from (\ref{iteration_10}) by neglecting all terms of
order $\OO4$. On the contrary, in the terms of the order of $\OO4$ in
(\ref{iteration_5}) and (\ref{iteration_10}) one can use the Newtonian
relation $\ve{\sigma} = \ve{k}$.

\subsection{The propagation time  $c\,\tau$}

Substituting (\ref{brumberg_40}) and (\ref{brumberg_50}) into
(\ref{iteration_5}) one can derive an explicit formula for the time of
light propagation as function of 
the given boundary conditions $\ve{x}_{\source}$ and $\ve{x}_{\obs}$:
\begin{eqnarray}
\fl {\phantom{\biggr|}}_{\rm N}&\biggr|& \quad 
c \, \tau = R
\nonumber\\
\fl {\phantom{\biggr|}}_{\rm pN}&\biggr|& \quad 
+ 
(1 + \gamma) \, m \, \log
\, \frac{x_{\obs} + x_{\source} + R}{x_{\obs} + x_{\source} - R}
\nonumber\\
\fl {\phantom{\biggr|}}_{\Delta\rm pN}&\biggr|& \quad 
+ \, \frac{1}{2}\,(1 + \gamma)^2 \, m^2 \, \frac{R}{|\ve{x}_{\obs} \times \ve{x}_{\source}|^2}
\, \left( (x_{\obs} - x_{\source})^2 - R^2 \right) 
\nonumber\\
\fl {\phantom{\biggr|}}_{\rm ppN}&\biggr|& \quad 
+ \, \frac{1}{8} \, \alpha \, \epsilon \, \frac{m^2}{R} \,
\left( \frac{x_{\source}^2 - x_{\obs}^2 - R^2}{x_{\obs}^2} \, + \,
\frac{x_{\obs}^2 - x_{\source}^2 - R^2}{x_{\source}^2} \right)
\nonumber\\
\fl {\phantom{\biggr|}}_{\rm ppN}&\biggr|&\quad 
+ \frac{1}{4} \, \alpha \, \left(8 (1 + \gamma) - 4 \beta + 3 \epsilon\right)\,
m^2 \,
\frac{R}{|\ve{x}_{\obs} \times \ve{x}_{\source}|} \;\delta \left(\ve{x}_{\obs} , \ve{x}_{\source}\right)
\nonumber\\
\fl && \quad +{\cal O}(c^{- 6}) \,.
\label{iteration_7}
\end{eqnarray}
\noindent
Here we have used that
$\delta (\ve{k} ,\ve{x}_{\source})-\delta(\ve{k},\ve{x}_{\obs})
=\delta(\ve{x}_{\obs},\ve{x}_{\source})$.
Here and below we classify the character of the individual terms by labels N
(Newtonian), pN (post-Newtonian), ppN (post-post-Newtonian) and
$\Delta\rm pN$ (terms that are formally of post-post-Newtonian order $\OO4$,
but may numerically become significantly larger than other post-post-Newtonian
terms; see below).  
Using $|\ve{x}_{\obs}\times\ve{x}_{\source}|=R\,d$ where $d$ is the impact parameter
defined by (\ref{impact_5}), 
and assuming general-relativistic
values of all parameters $\alpha=\beta=\gamma=\epsilon=1$ 
one gets the following estimates of the sums of the terms labelled 
by ``$\Delta\rm pN$'' and ``ppN'', respectively 
(the proofs are given in \cite{report2,report-proofs}):
\begin{eqnarray}
&&|c\,\delta\tau_{\Delta\rm pN}| 
\le2\,\frac{m^2}{d^2}\,R\,{4\,x_{\obs}\,x_{\source}\over (x_{\obs}+x_{\source})^2}
\,\le\,2\,\frac{m^2}{d^2}\,R\,,
\label{tau_20}
\\
&&|c\,\delta\tau_{\rm ppN}| \le \frac{15}{4}\;\pi\;\frac{m^2}{d}\,.
\label{tau_5}
\label{tau_25}
\end{eqnarray}
\noindent
These estimates and all estimates we give below
are reachable for some values of parameters and, in this sense, cannot
be improved. From these estimates we can conclude that among the
post-post-Newtonian terms $c\,\delta\tau_{\Delta\rm pN}$ can become
significantly larger compared to the other post-post-Newtonian terms.
For this reason we will call such terms ``enhanced''
post-post-Newtonian terms.  The physical origin and properties
of the ``enhanced'' post-post-Newtonian terms will be discussed
in Section \ref{section-physical-origin}.

The effect of $|c\,\delta\tau_{\rm ppN}|$ for the Sun is less than 3.7 cm for
arbitrary boundary conditions and can be neglected for any current and planned 
observations.
Therefore, the formula for the time of light propagation between two
given points can be simplified by taking only the relevant terms:
\begin{eqnarray}
\fl 
c \, \tau = R \, + \, (1 + \gamma)
\, m \, \log{\frac{x_{\obs} + x_{\source} + R}{x_{\obs} + x_{\source} - R}}
\nonumber
\\
\fl
\phantom{c\,\tau=}
- \, \frac{1}{2} \, (1 + \gamma)^2
\, m^2 \, \frac{R}{|\ve{x}_{\obs} \times \ve{x}_{\source}|^2}
\, \left(R^2- (x_{\obs} - x_{\source})^2 \right) +{\cal O}\left({m^2\over d}\right)+{\cal O}({m^3}).
\label{ctau-final}
\end{eqnarray}
\noindent
This expression can be written in an elegant form
\begin{eqnarray}
\fl c \, \tau = R\, + \, (1 + \gamma) \, m \,
\log{\frac{x_{\obs} + x_{\source} + R + (1 + \gamma) \, m}{x_{\obs} + x_{\source} - R+ (1 + \gamma) \, m}}
+{\cal O}\left({m^2\over d}\right)+{\cal O}({m^3})
\label{tau_30}
\end{eqnarray}
\noindent
that has been already derived in \cite{Moyer:2000} in an inconsistent
way (see Section 8.3.1.1 and Eq.~(8-54) of \cite{Moyer:2000}). As a
criterion if the additional post-post-Newtonian term is required for a given
situation, one can use (\ref{tau_20}) giving the upper boundary
of the additional term.

\subsection{Transformation from $\ve{k}$ to $\ve{\sigma}$}

Substituting (\ref{brumberg_40}) and (\ref{brumberg_50}) into
(\ref{iteration_10}) one gets:
\begin{eqnarray}
\fl {\phantom{\biggr|}}_{\rm N}&\biggr|& \quad {\ve{\sigma}} = {\ve{k}} 
\nonumber\\
\fl {\phantom{\biggr|}}_{\rm pN}&\biggr|& \quad
\phantom{\ve{\sigma}=} 
+ \, (1 + \gamma) \, m \, \frac{x_{\obs} - x_{\source} + R}
{|\ve{x}_{\obs} \times \ve{x}_{\source}|^2} \, {\ve{k}} \times (\ve{x}_{\source} \times \ve{x}_{\obs})
\nonumber\\
\fl {\phantom{\biggr|}}_{\Delta\rm pN}&\biggr|& \quad 
\phantom{\ve{\sigma}=} 
+\frac{\left(1+\gamma\right)^2}{2}\,m^2\,
{\ve{k}} \times \left( \ve{x}_{\source} \times \ve{x}_{\obs} \right)
\frac{\left(x_{\obs} + x_{\source}\right) 
\, (x_{\obs} - x_{\source} - R) (x_{\obs} - x_{\source} + R)^2}{|\ve{x}_{\obs} \times \ve{x}_{\source}|^4}  
\nonumber\\
\fl {\phantom{\biggr|}}_{\rm scaling}&\biggr|& \quad
\phantom{\ve{\sigma}=}  
- \, \frac{(1 + \gamma)^2}{2} \, m^2 \, \frac{(x_{\obs} - x_{\source} + R)^2}
{|\ve{x}_{\obs} \times \ve{x}_{\source}|^2} \, {\ve{k}}
\nonumber\\
\fl {\phantom{\biggr|}}_{\rm ppN}&\biggr|& \quad
\phantom{\ve{\sigma}=}  
 + \,m^2\,{\ve{k}} \times \left( \ve{x}_{\source} \times \ve{x}_{\obs} \right)\,
\Bigg[ - \frac{1}{4} \, \alpha \, \epsilon \,\frac{1}{R^2}\,
\left(\frac{1}{x_{\obs}^2}-\frac{1}{x_{\source}^2}\right)
\nonumber\\
\fl {\phantom{\biggr|}}_{\rm ppN}&\biggr|& \quad
\phantom{\ve{\sigma}=}  
+ \, \frac{1}{8} \, \left(8 (1 + \gamma - \alpha \, \gamma) (1 + \gamma)
- 4 \, \alpha \, \beta + 3 \alpha \, \epsilon\right)
\frac{1}{|\ve{x}_{\obs} \times \ve{x}_{\source}|^3}\,
\nonumber\\
\fl {\phantom{\biggr|}}_{\rm ppN}&\biggr|& \quad
\phantom{\ve{\sigma}=}  
\phantom{aaaaaaaaaa} \times\,
\biggl(2R^2 \left(\pi - \delta (\ve{k}, \ve{x}) \right)
+ \left(x_{\obs}^2 - x_{\source}^2- R^2\right) \delta (\ve{x}_{\obs} , \ve{x}_{\source}) \biggr)\Bigg]
\nonumber\\
\fl && \quad 
\phantom{\ve{\sigma}=} 
+{\cal O}(c^{- 6}) \,.
\label{k_to_sigma}
\end{eqnarray}
\noindent
This formula allows one to compute $\ve{\sigma}$ for given boundary
conditions $\ve{x}_{\source}$ and $\ve{x}_{\obs}$. Let us estimate the
magnitude of the individual terms in (\ref{k_to_sigma}) in the angle
$\delta(\ve{\sigma},\ve{k})$ between $\ve{\sigma}$ and $\ve{k}$. This
angle can be computed from vector product
$\ve{\rho}=\ve{k}\times\ve{\sigma}$, and, therefore, the term in
(\ref{k_to_sigma}) proportional to $\ve{k}$ and labelled as
``scaling'' plays no role.  Here and below terms proportional to
$\ve{k}$ do not influence the directions in the considered
approximation.  These terms are only necessary to keep the involved
vectors to have unit length. Now, we represent the vector product
$\ve{\rho}$ as the sum of three kinds of terms:
$\ve{\rho}=\ve{\rho}_{\rm pN}+\ve{\rho}_{\Delta\rm pN}+\ve{\rho}_{\rm
  ppN}$ where each term is the vector product of $\ve{k}$ and the
sum of the correspondingly labelled terms in (\ref{k_to_sigma}).  Using
\begin{eqnarray}
\left|\, \ve{k} \times \left[ \ve{k} \times
\left( \ve{x}_{\source} \times \ve{x}_{\obs} \right) \right]  \,\right| &=&
\left|\, \ve{k} \times \left( \ve{x}_{\source} \times \ve{x}_{\obs} \right) \,\right|
\,=\, R\, d \,,
\label{sigma_10}
\end{eqnarray}
\noindent
and general-relativistic values of the parameters
$\alpha=\beta=\gamma=\epsilon=1$ one
gets (the proofs can be found in
\cite{report2,report-proofs}):
\begin{eqnarray}
&&|\ve{\rho}_{\rm pN}| \le
{4m\over d}\,
\left[
\  
\begin{array}{ll}
1,\quad&  x_{\source}\le x_{\obs}, \\[5pt] 
\displaystyle{\frac{x_{\obs}}{x_{\obs} + x_{\source}}}, & x_{\source}>x_{\obs}
\end{array}
\right.\,
\le {4m\over d}\,,
\label{estimate-rho-0}
\\
&&|\ve{\rho}_{\Delta\rm pN}| \le 16\, \frac{m^2}{d^3} 
\left[
\  
\begin{array}{ll}
\displaystyle{4\over 27}\,(x_{\obs}+x_{\source}),\quad&  \displaystyle{1\over 2}
\,x_{\obs}\le x_{\source}\le x_{\obs}, \\[5pt] 
\displaystyle{\frac{x_{\obs}^2\,x_{\source}}{\left(x_{\obs} + x_{\source}\right)^2}}, 
& x_{\source}<{1\over 2}\,x_{\obs}\ {\rm or}\ x_{\source}>x_{\obs}\,,
\end{array}
\right.
\label{rho-4} 
\\
&&|\ve{\rho}_{\rm ppN}| \le  
\frac{15}{4}\,\pi\,\frac{m^2}{d^2}\,.
\label{sigma_30}
\label{sigma_40}
\end{eqnarray}
\noindent
Note that $\ve{\rho}_{\rm pN}$ and $\ve{\rho}_{\Delta\rm pN}$
themselves as well as their estimates are not continuous for
$\ve{x}_{\obs}\to\ve{x}_{\source}$ since in this limit an infinitely small change of
$\ve{x}_{\obs}$ leads to big changes in $\ve{k}$. Discontinuity of the same
origin appears for many other terms. The limit $\ve{x}_{\obs}\to\ve{x}_{\source}$ and
the corresponding discontinuity have, clearly, no physical importance.

We see that among terms of order ${\cal O}(m^2)$ only 
$|\ve{\rho}_{\Delta\rm  pN}|$ cannot be estimated as ${\rm const}\times m^2/d^2$.  
The sum of the three other terms labelled as ``ppN'' can be estimated as given by 
(\ref{sigma_30}).
In most cases these terms can be neglected at the
level of 1 \muas. Indeed, it is easy to see that $|\ve{\rho}_{\rm ppN}|$ can 
exceed 1 \muas\ only for observations within
about 3.3 angular radii from the Sun.
Accordingly, we obtain a simplified
formula for the transformation from $\ve{k}$ to $\ve{\sigma}$ keeping 
only the post-Newtonian and ``enhanced'' post-post-Newtonian terms labelled
as ``pN'' and ``$\Delta$pN'' in (\ref{k_to_sigma}):
\begin{eqnarray}
\fl \ve{\sigma} = \ve{k}\,+\ve{d}\,S\,
\left(1-S\,{1\over 2}\,(x_{\obs}+x_{\source})\left(1+{x_{\source}-x_{\obs}\over R}\right)\right)
+{\cal O}\left({m^2\over d^2}\right)+{\cal O}({m^3})
\,,
\label{sigma-k-better}
\\
\fl S = (1+\gamma) {m\over d^2} \left(1-{x_{\source}-x_{\obs}\over R}\right) \,.
\label{sigma-k-S}
\end{eqnarray}
\noindent
Eq. (\ref{rho-4}) can be used as a criterion if the 
post-post-Newtonian term in (\ref{sigma-k-better}) is necessary for a
given accuracy and configuration.

\subsection{Transformation from $\ve{\sigma}$ to $\ve{n}$}

The transformation between $\ve{n}$ and $\ve{\sigma}$ is given by
(\ref{sigma2n-formal})--(\ref{n_5}).  We need, however, to express the
relativistic terms in (\ref{sigma2n-formal}) as functions of $\ve{k}$.
To this end we note that $\ve{x}_{\rm pN}=\ve{x}_{\obs}+{\cal O}(c^{-
  4})$ and $\ve{x}_{\obs}=\ve{x}_{\source}+R\,\ve{k}$, use
(\ref{k_to_sigma}) for $\ve{\sigma}$ in $\ve{C}_1(\ve{x}_{\rm
  pN})$, and get
\begin{eqnarray}
\fl {\phantom{\biggr|}}_{\rm N}&\biggr|&\quad 
\ve{n} =  \ve{\sigma}
\nonumber\\
\fl {\phantom{\biggr|}}_{\rm pN}&\biggr|&\quad 
-(1 + \gamma)\,m\,
\ve{k} \times ( \ve{x}_{\source} \times \ve{x}_{\obs} )
\frac{R}{|\, \ve{x}_{\obs} \times \ve{x}_{\source} \,|^2}\,
\left(1\,+\,\frac{\ve{k} \cdot \ve{x}_{\obs}}{x_{\obs}}\right)
\nonumber\\
\fl {\phantom{\biggr|}}_{\rm scaling}&\biggr|&\quad
+\frac{1}{4}\,(1 + \gamma)^2\,m^2\,
\frac{\ve{k}}{|\,\ve{x}_{\obs} \times \ve{x}_{\source}\,|^2}\,
{R\over x_{\obs}}\,\left(1\,+\,\frac{\ve{k} \cdot \ve{x}_{\obs}}{x_{\obs}}\right)\,
(3x_{\obs} - x_{\source} - R)\,(x_{\obs} - x_{\source} + R)
\nonumber\\
\fl {\phantom{\biggr|}}_{\Delta\rm pN}&\biggr|&\quad
 + m^2\,\ve{k} \times ( \ve{x}_{\source} \times \ve{x}_{\obs} ) \,
\Bigg[ \,(1 + \gamma)^2\,\frac{x_{\obs} + x_{\source}}{|\,\ve{x}_{\obs} \times \ve{x}_{\source}\,|^2}
\,\left( 1 + \frac{\ve{k} \cdot \ve{x}_{\obs}}{x_{\obs}} \right)
\frac{R\,\left(R^2-(x_{\obs}-x_{\source})^2\right)}{2\,|\,\ve{x}_{\obs} \times \ve{x}_{\source}\,|^2}
\nonumber\\
\fl {\phantom{\biggr|}}_{\rm ppN}&\biggr|&\quad
\qquad
+(1 + \gamma)^2\,\frac{R}{|\,\ve{x}_{\obs} \times \ve{x}_{\source}\,|^2}
\,\left( 1 + \frac{\ve{k} \cdot \ve{x}_{\obs}}{x_{\obs}} \right)\,
{1\over x_{\obs}}
\nonumber\\
\fl {\phantom{\biggr|}}_{\rm ppN}&\biggr|&\quad
\qquad + \frac{1}{2}\,(1 + \gamma)^2\,\frac{R^2}{|\,\ve{x}_{\obs} \times \ve{x}_{\source}\,|^4}
\,\left( 1 + \frac{\ve{k} \cdot \ve{x}_{\obs}}{x_{\obs}} \right)\left(1-\frac{x_{\obs}+x_{\source}}{R}\right)
\left(R^2-(x_{\obs}-x_{\source})^2\right)
\nonumber\\
\fl {\phantom{\biggr|}}_{\rm ppN}&\biggr|&\quad
\qquad
-\frac{1}{2}\,\alpha\,\epsilon\,\frac{\ve{k} \cdot \ve{x}_{\obs}}{R\,x_{\obs}^4}
\nonumber\\
\fl {\phantom{\biggr|}}_{\rm ppN}&\biggr|&\quad
\qquad
- {1\over 4}\,\left(8\,(1 + \gamma - \alpha\,\gamma)(1 + \gamma)
\,-4\,\alpha\,\beta\,+3\,\alpha\,\epsilon \right)
\frac{\ve{k} \cdot \ve{x}_{\obs}}{x_{\obs}^2}\,\frac{R}{|\,\ve{x}_{\obs} \times \ve{x}_{\source}\,|^2}
\nonumber\\
\fl {\phantom{\biggr|}}_{\rm ppN}&\biggr|&\quad
\qquad
- {1\over 4}\,
\left(8 (1 + \gamma-\alpha\,\gamma)(1 + \gamma)\,-4\,\alpha\,\beta\,
+ 3\,\alpha\,\epsilon \right)\,\frac{R^2}{|\,\ve{x}_{\obs} \times \ve{x}_{\source}\,|^3}\,
\left( \pi - \delta (\ve{k} , \ve{x}_{\obs}) \right) 
\, \Bigg] 
\nonumber\\
\fl && \quad + {\cal O}(c^{- 6})\,.
\label{n_10}
\end{eqnarray}
\noindent
This expression allows one to compute the difference between the
vectors $\ve{n}$ and $\ve{\sigma}$ starting from the boundary
conditions $\ve{x}_{\source}$ and $\ve{x}_{\obs}$.  Let us estimate
the magnitude of the individual terms in (\ref{n_10}) in the angle
$\delta(\ve{\sigma},\ve{n})$ between $\ve{n}$ and $\ve{\sigma}$. This
angle can be computed from vector product
$\ve{\varphi}=\ve{\sigma}\times\ve{n}$.  Again the term in
(\ref{n_10}) proportional to $\ve{k}$ and labelled as ``scaling''
plays no role since $\ve{\sigma}\times\ve{k}=\OO2$. In order to
estimate the effects of the other terms in (\ref{n_10}), we split
$\ve{\varphi}=\ve{\varphi}_{\rm pN}+\ve{\varphi}_{\Delta\rm pN}+\ve{\varphi}_{\rm
  ppN}$ similarly as we did with $\ve{\rho}$ above, take into
account that $|\,\ve{\sigma} \times \bigl( \ve{k} \times (
\ve{x}_{\source} \times \ve{x}_{\obs}) \bigr)\,| = R\, d + \OO2$,
assume again $\alpha = \beta = \gamma = \epsilon = 1$ and get
\cite{report2,report-proofs}:
\begin{eqnarray}
\fl |\ve{\varphi}_{\rm pN}| = 2\,m\,
\biggl|\,\ve{\sigma} \times [ \ve{k} \times ( \ve{x}_{\source} \times \ve{x}_{\obs}) ] 
\biggr|\,
\frac{R}{|\,\ve{x}_{\obs} \times \ve{x}_{\source}\,|^2}
\left(1 + \frac{\ve{k} \cdot \ve{x}_{\obs}}{x_{\obs}}\right)
 \,\le\,4\,\frac{m}{d}\,,
\label{estim_5}
\\
\fl |\ve{\varphi}_{\Delta\rm pN}| = 4\,m^2\,
\biggl|\,\ve{\sigma} \times [ \ve{k} \times ( \ve{x}_{\source} \times \ve{x}_{\obs}) ] \,
\biggr|\,
\left(1 + \frac{\ve{k} \cdot \ve{x}_{\obs}}{x_{\obs}}\right)\,
\frac{R\,\left(x_{\obs}+x_{\source}\right)}{|\,\ve{x}_{\obs} \times \ve{x}_{\source}\,|^4}\,
\frac{R^2 - (x_{\obs} - x_{\source})^2}{2}
\nonumber\\
\fl 
\phantom{|\ve{\varphi}_{\Delta\rm pN}| =}
\le 4\,\frac{m^2}{d^2}\,\frac{4x_{\obs}\,x_{\source}}{d\,(x_{\obs} + x_{\source})}
\le 16\,\frac{m^2}{d^2}\,{x_{\obs}\over d},
\label{estim_10}
\\
\fl 
|\ve{\varphi}_{\rm ppN}| \le {15\over 4}\,\pi\,{m^2\over d^2}.
\label{estim_30}
\label{n_20}
\end{eqnarray}
\noindent 
Eq. (\ref{estim_30}) shows that the ``ppN'' terms can attain 1 \muas\
only if one observes within approximately 3.3 angular radii from the
Sun. In many cases these terms can be neglected.  Accordingly, we
obtain a simplified formula for the transformation from $\ve{\sigma}$
to $\ve{n}$ keeping only the post-Newtonian and ``enhanced''
post-post-Newtonian terms labelled as ``pN'' and ``$\Delta$pN'' in
(\ref{n_10}):
\begin{eqnarray}
\ve{n} &=& \ve{\sigma}+\ve{d}\,T\,\left(1+T\,x_{\obs}\,
{R+x_{\source}-x_{\obs}\over R+x_{\source}+x_{\obs}}
\right)
+{\cal O}\left({m^2\over d^2}\right)+{\cal O}({m^3})\,,
\label{n-sigma-better}
\\
T&=& -(1 + \gamma)\,\frac{m}{d^2}\,
\left(1 + \frac{\ve{k} \cdot \ve{x_{\obs}}}{x_{\obs}}\right)\,.
\label{n-sigma-T}
\end{eqnarray}
\noindent
Eq. (\ref{estim_10}) can be used as a criterion if the additional
post-post-Newtonian term in (\ref{n-sigma-better}) is necessary for a
given accuracy and configuration.

\subsection{Transformation from $\ve{k}$ to $\ve{n}$}

Finally, a direct relation between vectors $\ve{k}$ and $\ve{n}$
should be derived. To this end, we combine
(\ref{k_to_sigma}) and (\ref{n_10}) to get
\begin{eqnarray}
\fl {\phantom{\biggr|}}_{\rm N}&\biggr|&\quad 
\ve{n} = \ve{k}
\nonumber\\
\fl  {\phantom{\biggr|}}_{\rm pN}&\biggr|&\quad 
\phantom{\ve{n} = \ve{k}}
 - (1 + \gamma) \, m \,\frac{\ve{k} \times
(\ve{x}_{\source} \times \ve{x}_{\obs})}{x_{\obs}\left(x_{\obs}\,x_{\source} + \ve{x}_{\obs}\cdot\ve{x}_{\source}\right)}
\nonumber\\
\fl {\phantom{\biggr|}}_{\Delta\rm pN}&\biggr|&\quad
\phantom{\ve{n} = \ve{k}} 
 + (1 + \gamma)^2 \, m^2 \,\frac{\ve{k} \times
(\ve{x}_{\source} \times \ve{x}_{\obs})}{\left(x_{\obs}\,x_{\source} + \ve{x}_{\obs}\cdot\ve{x}_{\source}\right)^2}\,{x_{\obs} + x_{\source}\over x_{\obs}}
\nonumber\\
\fl {\phantom{\biggr|}}_{\rm scaling}&\biggr|&\quad
\phantom{\ve{n} = \ve{k}} 
- \frac{1}{8}\,(1 + \gamma)^2\,\frac{m^2}{x_{\obs}^2}\,\ve{k}\,
\frac{{\left((x_{\obs} - x_{\source})^2 - R^2\right)}^2}{|\ve{x}_{\obs} \times \ve{x}_{\source}|^2}
\nonumber\\
\fl {\phantom{\biggr|}}_{\rm ppN}&\biggr|&\quad
\phantom{\ve{n} = \ve{k}} 
 + \,m^2\, \ve{k} \times (\ve{x}_{\source} \times \ve{x}_{\obs})\,
\Biggl[
\,{1\over 2}\,(1 + \gamma)^2\,
\frac{R^2-(x_{\obs}-x_{\source})^2}{x_{\obs}^2\,|\ve{x}_{\obs} \times \ve{x}_{\source}|^2}
\nonumber\\
\fl {\phantom{\biggr|}}_{\rm ppN}&\biggr|&\quad
\phantom{\ve{n} = \ve{k}} 
 + \, \frac{1}{4} \, \alpha \, \epsilon \, \frac{1}{R}
\left(\frac{1}{R\,x_{\source}^2} - \frac{1}{R\,x_{\obs}^2}
- 2\, \frac{\ve{k} \cdot \ve{x_{\obs}}}{x_{\obs}^4}\right)
\nonumber\\
\fl {\phantom{\biggr|}}_{\rm ppN}&\biggr|&\quad
\phantom{\ve{n} = \ve{k}} 
 - \frac{1}{4}\,\left(\, 8(1 + \gamma - \alpha \gamma) (1 + \gamma) - 4\alpha \beta
+ 3\, \alpha\, \epsilon \, \right)  \, R\,\frac{\ve{k} \cdot \ve{x}_{\obs}}
{x_{\obs}^2\,|\, \ve{x}_{\obs} \times \ve{x}_{\source} \,|^2}
\nonumber\\
\fl {\phantom{\biggr|}}_{\rm ppN}&\biggr|&\quad
\phantom{\ve{n} = \ve{k}} 
 + \frac{1}{8}\, \left(8 (1 + \gamma - \alpha \, \gamma) (1 + \gamma)
- 4 \, \alpha \, \beta + 3 \alpha \, \epsilon\right) \,
\frac{x_{\obs}^2 - x_{\source}^2 - R^2}{|{\ve{x}_{\obs}} \times {\ve{x}}_{\source}|^3} 
\,\delta(\ve{x}_{\obs} , \ve{x}_{\source})
\Biggr]
\nonumber\\
\fl && \quad  
\phantom{\ve{n} = \ve{k}}
+\OO6\,.
\label{n_60}
\end{eqnarray}
\noindent
This formula allows one to compute the unit coordinate direction
of light propagation $\ve{n}$ at the point of reception starting from
the positions of the source $\ve{x}_{\source}$ and the observer $\ve{x}_{\obs}$.

As in other cases our goal now is to estimate the effect of the
individual terms in (\ref{n_60}) on the angle $\delta(\ve{k},\ve{n})$
between $\ve{k}$ and $\ve{n}$. This angle can be computed from vector
product $\ve{\omega}=\ve{k}\times\ve{n}$.  The term in (\ref{n_60})
proportional to $\ve{k}$ and labelled by ``scaling'' obviously plays
no role here and can be ignored. For the other terms in
$\ve{\omega}=\ve{\omega}_{\rm pN}+\ve{\omega}_{\Delta\rm
  pN}+\ve{\omega}_{\rm ppN}$ taking into account (\ref{sigma_10}) and
considering the general-relativistic values
$\alpha=\beta=\gamma=\epsilon=1$ one gets \cite{report2,report-proofs}
\begin{eqnarray}
|\ve{\omega}_{\rm pN}| &=& 2\,m \,\frac{1}{x_{\obs}}\,
\, \frac{\left|\ve{k} \times (\ve{x}_{\source} \times \ve{x}_{\obs})\right|}
{x_{\obs}\,x_{\source} + \ve{x}_{\obs} \cdot \ve{x}_{\source}}
\,\le\, 4\,\frac{m}{d}\,\frac{x_{\source}}{x_{\obs} \, + \,x_{\source}}\,\le\, 4\,\frac{m}{d}\,,
\label{omega_5}
\\
\nonumber\\
|\ve{\omega}_{\Delta\rm pN}| &=& 4\,m^2\,\frac{x_\obs+x_\source}{x_\obs}\,
\frac{\left|\ve{k} \times (\ve{x}_{\source} \times \ve{x}_{\obs})\right|}
{\left(x_{\obs}\,x_{\source} + \ve{x}_{\obs} \cdot \ve{x}_{\source}\right)^2}
\nonumber\\
&\le& 
16\,\frac{m^2}{d^3}\,\frac{R\,x_{\obs}\,x_{\source}^2}{(x_{\obs}+x_{\source})^3}
\,\le\, 16\,\frac{m^2}{d^3}\,\frac{x_{\obs}\,x_{\source}^2}{(x_{\obs}+x_{\source})^2} 
\le 16 {m^2\over d^2}\, {x_{\obs}\over d}\,,
\label{omega-1}
\end{eqnarray}
\noindent
or, alternatively,
\begin{eqnarray}
|\ve{\omega}_{\Delta\rm pN}| &\le&  
{64\over 27}\,\frac{m^2}{d^2}\,{R\over d}\,.
\label{omega-1-alternative}
\end{eqnarray}
\noindent
We give four possible estimates of $|\ve{\omega}_{\Delta\rm pN}|$. 
These estimates can be useful in different situations.
Note that the last estimate in (\ref{omega-1})
and the estimate in (\ref{omega-1-alternative})
cannot be related to each other and reflect different properties of 
$|\ve{\omega}_{\Delta\rm pN}|$ as function of multiple variables.

The effect of all the ``ppN'' terms in (\ref{n_60}) can be estimated as
\begin{eqnarray}
|\ve{\omega}_{\rm ppN}| \le
{15\over 4}\,\pi\, {m^2\over d^2}\,.
\label{omega_30}
\label{omega_22}
\end{eqnarray}
Again these terms can attain 1 \muas\ only for observations within
about 3.3 angular radii from the Sun and can be
neglected. Accordingly, we obtain a simplified formula for the
transformation from $\ve{k}$ to $\ve{n}$ keeping only the
post-Newtonian and ``enhanced'' post-post-Newtonian terms labelled as
``pN'' and ``$\Delta$pN'' in (\ref{n_60}):
\begin{eqnarray}
\ve{n} &=& \ve{k} 
+\ve{d}\,P\,\left(1+P\,x_{\obs}\,{x_{\source}+x_{\obs}\over R}\right)
+{\cal O}\left({m^2\over d^2}\right)+{\cal O}({m^3})\,,
\label{n_85-better}
\\
P&=&-(1+\gamma)\,{m\over d^2}\,
\left({x_{\source}-x_{\obs}\over R}+{\ve{k}\cdot\ve{x}_{\obs}\over x_{\obs}}\right)\,.
\label{P-sso}
\end{eqnarray}
\noindent
Let us also note that the post-post-Newtonian term in
(\ref{n_85-better}) is maximal for sources at infinity:
\begin{eqnarray}
 |\ve{\omega}_{\Delta\rm pN}| \le \lim_{x_{\source}\to\infty} |\ve{\omega}_{\Delta\rm pN}| 
&=& \lim_{x_{\source} \to \infty} (1 + \gamma)^2\,m^2\,
\frac{x_{\source}+x_{\obs}}{x_{\obs}}\,
\frac{\left|\ve{k} \times (\ve{x}_{\source} \times \ve{x}_{\obs})\right|}
{\left(x_{\obs}\,x_{\source} + \ve{x}_{\obs} \cdot \ve{x}_{\source}\right)^2}
\nonumber\\
&=& (1 + \gamma)^2\,
(1-\cos\Phi)^2\,\frac{m^2}{d^2}\,{x_{\obs}\over d}\,,
\label{n_95}
\end{eqnarray}

\noindent
where $\Phi=\delta(\ve{x}_{\source},\ve{x}_{\obs})$ is the angle between vectors
$\ve{x}_{\source}$ and $\ve{x}_{\obs}$. 
Several useful estimates of this term are
given by (\ref{omega-1})--(\ref{omega-1-alternative}). These estimates
can be used as a criterion which allows one to decide if
the post-post-Newtonian correction is important for a particular
situation.

\subsection{Transformation from $\ve{k}$ to $\ve{n}$ for  stars and quasars}
\label{section-stars}

In principle, the formulas for the boundary problem given above are valid
also for stars and quasars. However, for sufficiently large $x_{\source}$
the formulas could be simplified. It is the purpose of this
Section to derive the formulas for this case.

\subsubsection{Transformation from $\ve{k}$ to $\ve{\sigma}$.}

First, let us show that for stars and quasars the approximation
\begin{eqnarray}
\ve{\sigma} &=& \ve{k}\,
\label{stars_5}
\end{eqnarray}

\noindent
is valid for an accuracy of 1 \muas.  Using estimates
(\ref{estimate-rho-0}) and (\ref{rho-4}) for the two terms in
(\ref{sigma-k-better}) one can see that for $x_{\source}\gg x_{\obs}$ the angle
$\delta(\ve{\sigma},\ve{k})$ can be estimated as
\begin{equation}
\delta(\ve{\sigma},\ve{k}) \le 4\, {m\over d}\,{x_{\obs}\over x_{\obs}+x_{\source}}\,
\left(1+ 4\,{m\over d}\,{x_{\obs}\over d}\,{x_{\source}\over x_{\obs}+x_{\source}}\,\right).
\label{sigma-k-stars-estimates}
\end{equation}

\noindent
Clearly, $\delta(\ve{\sigma},\ve{k})$ goes to zero for
$x_{\source}\to\infty$.  Numerical values of this upper estimate are
given in Table \ref{table_stars} for $x_{\source}$ equal to 1, 10 and
100 pc.  Angle $\delta(\ve{\sigma},\ve{k})$ is smaller for stars at
larger distances. However, for objects with $x_{\source} < 1$ pc the
difference between $\ve{\sigma}$ and $\ve{k}$ must be explicitly taken
into account. From the point of view of the relativistic model these
objects should be treated in the same way as solar system objects.

\begin{table}
\caption{
\label{table_stars}
Numerical values of estimate (\ref{sigma-k-stars-estimates})
in \muas\ for the angle between $\ve{\sigma}$ and $\ve{k}$ due to the solar system bodies
for various values of $x_{\source}$.}
\begin{indented}
\item[]
\begin{tabular}{@{}rllllll}
\br
$x_{\source}$ [pc]&Sun&Sun at $45^\circ$&Jupiter&Saturn&Uranus&Neptune\\[3pt]
\mr
&&&&&\\[-10pt]
1   & 8.506 & 0.056 & 0.473 & 0.309 & 0.212 & 0.382 \\
10  & 0.851 & 0.006 & 0.047 & 0.031 & 0.021 & 0.038 \\
100 & 0.085 & 0.001 & 0.004 & 0.003 & 0.002 & 0.004 \\
\br
\end{tabular}
\end{indented}
\end{table}

\subsubsection{Transformation from $\ve{\sigma}$ to $\ve{n}$.}

As soon as we accept the equality of $\ve{\sigma}$ and $\ve{k}$ for
stars the only relevant step is  
the transformation between $\ve{\sigma}$ and $\ve{n}$. 
This transformation in the post-post-Newtonian
approximation is given by (\ref{sigma2n-formal})--(\ref{n_5}).
In the framework of the relativistic light deflection model,
the distances to stars and quasars are assumed to be unknown and so large
that they can be considered infinitely large. 
For such sources it is natural to use the observer's position
$\ve{x}_\obs$ as initial position denoted in 
(\ref{cauchy_5}) as $\ve{x}_0$. Therefore, in 
(\ref{impact_10}) and (\ref{impact_25}) one should formally
replace $\ve{x}_0$ by $\ve{x}_\obs$. E.g., the impact
parameter $\ve{d}_\sigma$ is defined as
\begin{equation}
\ve{d}_\sigma=\ve{\sigma}\times(\ve{x}_{\obs}\times\ve{\sigma}).
\end{equation}
\noindent
We can rewrite (\ref{sigma2n-formal})--(\ref{n_5}) as
\begin{eqnarray}
\fl {\phantom{\biggr|}}_{\rm N}&\biggr|&\quad 
\ve{n}=\ve{\sigma} 
\nonumber\\
\fl {\phantom{\biggr|}}_{\rm pN}&\biggr|&\quad
\phantom{\ve{n}=\ve{\sigma}}  
- (1+\gamma)\,m\,{\ve{d}_\sigma\over d_\sigma^2}\,
\left(1+{\ve{\sigma}\cdot\ve{x}_{\obs}\over x_{\obs}}\right)
\nonumber\\
\fl {\phantom{\biggr|}}_{\Delta\rm pN}&\biggr|&\quad
\phantom{\ve{n}=\ve{\sigma}}   
+(1 + \gamma)^2\,m^2
\frac{\ve{d}_\sigma}{d_\sigma^3}\,{x_{\obs}\over d_\sigma}
{\left(1+{\ve{\sigma}\cdot\ve{x}_{\obs}\over x_{\obs}}\right)}^2 
\nonumber\\
\fl {\phantom{\biggr|}}_{\rm scaling}&\biggr|&\quad
\phantom{\ve{n}=\ve{\sigma}}   
-\frac{1}{2}\,m^2
(1 + \gamma)^2\,\frac{\ve{\sigma}}{d_\sigma^2}
{\left(1+{\ve{\sigma}\cdot\ve{x}_{\obs} \over x_{\obs}}\right)}^2
\nonumber\\
\fl {\phantom{\biggr|}}_{\rm ppN}&\biggr|&\quad
\phantom{\ve{n}=\ve{\sigma}}   
- \frac{1}{2}\,m^2\alpha\,\epsilon\, \frac{\ve{\sigma} \cdot \ve{x}_{\obs}}{x_{\obs}^4}\,
\ve{d}_\sigma
\nonumber\\
\fl {\phantom{\biggr|}}_{\rm ppN}&\biggr|&\quad
\phantom{\ve{n}=\ve{\sigma}}   
+(1 + \gamma)^2\,m^2\frac{\ve{d}_\sigma}{d_\sigma^2}\,
{1\over x_{\obs}}\,\left(1+{\ve{\sigma}\cdot\ve{x}_{\obs}\over x_{\obs}}\right)
\nonumber\\
\fl {\phantom{\biggr|}}_{\rm ppN}&\biggr|&\quad
\phantom{\ve{n}=\ve{\sigma}}   
-{1\over 4}\,\left(8\,(1+\gamma-\alpha\,\gamma)\,(1+\gamma)-4\,\alpha\,\beta+3\,\alpha\,\epsilon\right)\,m^2\,
{\ve{d}_\sigma\over d_\sigma^2}\,
{\ve{\sigma}\cdot\ve{x}_{\obs}\over x_{\obs}^2}
\nonumber\\
\fl {\phantom{\biggr|}}_{\rm ppN}&\biggr|&\quad
\phantom{\ve{n}=\ve{\sigma}}   
-{1\over 4}\,\left(8\,(1+\gamma-\alpha\,\gamma)\,(1+\gamma)-4\,\alpha\,\beta+3\,\alpha\,\epsilon\right)\,m^2\,
{\ve{d}_\sigma\over d_\sigma^3}\,
\left( \pi - \delta (\ve{\sigma} , \ve{x}_{\obs}) \right)
\nonumber\\
\fl && \quad 
\phantom{\ve{n}=\ve{\sigma}}  
+ {\cal O}(m^3),
\label{sigma-n-stars}
\end{eqnarray}
\noindent
where $d_\sigma=|\ve{d}_\sigma|=|\ve{\sigma}\times\ve{x}_{\obs}|$. Now
we need to estimate the effect of the individual terms in
(\ref{sigma-n-stars}) on the angle $\delta(\ve{\sigma},\ve{n})$
between $\ve{\sigma}$ and $\ve{n}$. This angle can be computed from
vector product $\ve{\psi}=\ve{\sigma}\times\ve{n}$.  The term in
(\ref{sigma-n-stars}) proportional to $\ve{\sigma}$ and labelled as
``scaling'' obviously plays no role and can be ignored.  For the other
terms in $\ve{\psi}=\ve{\psi}_{\rm pN}+\ve{\psi}_{\Delta\rm
  pN}+\ve{\psi}_{\rm ppN}$ taking into account that
$|\ve{\sigma}\times\ve{d}_\sigma|=d_\sigma$ and considering the
general-relativistic values $\alpha=\beta=\gamma=\epsilon=1$ we get
\cite{report2,report-proofs}
\begin{eqnarray}
&&|\ve{\psi}_{\rm pN}|=
2\,m\,{|\ve{\sigma}\times\ve{d}_\sigma|\over d_\sigma^2}\,
\left(1+{\ve{\sigma}\cdot\ve{x}_{\obs}\over x_{\obs}}\right)
\le4\,{m\over d_\sigma}\,,
\label{psi-0}
\\
&&|\ve{\psi}_{\Delta\rm pN}|=4m^2\,
\frac{|\ve{\sigma}\times\ve{d}_\sigma|}{d_\sigma^3}\,{x_{\obs}\over d_\sigma}
{\left(1+{\ve{\sigma}\cdot\ve{x}_{\obs}\over x_{\obs}}\right)}^2 
\le 16\,{m^2\over d_\sigma^2}\,{x_{\obs}\over d_\sigma}\,,
\label{psi-3}
\\
&&|\ve{\psi}_{\rm ppN}|\le
{15\over 4}\,\pi\,{m^2\over d_\sigma^2}\,.
\label{psi-5}
\label{psi-estimate}
\end{eqnarray}

\noindent
The estimate shows that the ``ppN'' terms can be neglected at the level of 
1 \muas\ except for the observations within about 3.3 
angular radii from the Sun. Omitting these terms one gets
an expression valid at the level of 1 \muas\ in all other cases:
\begin{eqnarray}
\ve{n}&=&\ve{\sigma}+\ve{d}_\sigma\,Q\,(1+Q\,x_{\obs}) 
+ {\cal O}\left({m^2\over d_\sigma^2}\right)
+ {\cal O}(m^3)\,,
\label{sigma-n-stars-simplified}
\\
Q&=&- (1+\gamma)\,{m\over d_\sigma^2}\,
\left(1+{\ve{\sigma}\cdot\ve{x}_{\obs}\over x_{\obs}}\right)\,.
\label{Q-stars}
\end{eqnarray}

\noindent
This coincides with (\ref{n_85-better})--(\ref{P-sso}) and with
(\ref{n-sigma-better})--(\ref{n-sigma-T}) for $x_{\source}\to\infty$.
This formula together with $\ve{\sigma}=\ve{k}$ can be applied for
sources at distances larger than 1 pc to attain the accuracy of 1
\muas. Alternatively, Eqs. (\ref{n_85-better})--(\ref{P-sso}) can be
used for the same purpose giving slightly better accuracy for very
close stars. However, distance information (parallax) is necessary to
use (\ref{n_85-better})--(\ref{P-sso}).

\subsection{Numerical estimates and Monte-Carlo simulations}
\label{section-numerical-estimates}

Table~\ref{table2} contains numerical values of the ``regular'' 
post-post-Newtonian terms of order 
${\cal O}(m^2/d)$ in (\ref{iteration_7}) and
of order ${\cal O}(m^2/d^2)$ in (\ref{k_to_sigma}),
(\ref{n_10}), (\ref{n_60}), and (\ref{sigma-n-stars}).
The analytical estimates are given by (\ref{tau_25}),
(\ref{sigma_40}), (\ref{n_20}),  (\ref{omega_22}), and 
(\ref{psi-estimate}), respectively. One can see that at the 
level of 10 cm in distances and 1 \muas\ in angles
these terms are irrelevant except for observations within 
3.3 angular radii from the Sun. 

A series of additional Monte-Carlo simulations using randomly chosen
boundary conditions has been performed to verify the given estimates
of the post-post-Newtonian terms numerically. The results of these simulations
fully confirm all our estimates.

\begin{table}
\caption{
\label{table2}
Numerical values of the analytical 
upper estimates of the post-post-Newtonian
terms of order of ${\cal O}(m^2/d)$ in (\ref{iteration_7}) and
of order ${\cal O}(m^2/d^2)$ in (\ref{k_to_sigma}),
(\ref{n_10}), (\ref{n_60}), and (\ref{sigma-n-stars}).
}
\begin{indented}
\item[]
\begin{tabular}{@{}rllllll}
\mr
&Sun&Sun at $45^\circ$&Jupiter&Saturn&Uranus&Neptune\\
\mr
$|c\,\delta\tau_{\rm ppN}|$ \ [$10^{-6}$ m] & 36906.0 & 242.9 & 0.328 & 0.036 & 0.002 & 0.003 \\
$\begin{array}{l}
|\ve{\rho}_{\rm ppN}|,\ 
|\ve{\varphi}_{\rm ppN}|,\\ 
|\ve{\omega}_{\rm ppN}|,\ 
|\ve{\psi}_{\rm ppN}|
\end{array}$
\ [$10^{-3}$\ \muas] & 10937.4 & 0.474 & 0.945 & 0.120 & 0.016 & 0.023\\
\br
\end{tabular}
\end{indented}
\end{table}

\begin{table}
\caption{
\label{table5}
Maximal numerical value (\ref{n_95})
of the ``enhanced'' post-post-Newtonian term 
in (\ref{n_85-better}) for the solar system bodies with parameters 
given in Table \ref{table0}.
}
\begin{indented}
\item[]
\begin{tabular}{@{}rllllll}
\br
&Sun&Sun at $45^{\circ}$&Jupiter&Saturn&Uranus&Neptune\\
\mr
$\max|\ve{\omega}_{\Delta\rm pN}|$\ [\muas]
&$3192.8$&$0.663 \times 10^{-3}$&$16.11$&$4.42$&$2.58$&$5.83$\\
\br
\end{tabular}
\end{indented}
\end{table}

Using estimate (\ref{n_95}) and the parameters of the solar system
bodies given in Table \ref{table0} one can compute the maximal values
of the ``enhanced'' post-post-Newtonian term in the transformation
from $\ve{k}$ to $\ve{n}$.  For grazing rays one can apply $\cos\Phi
\simeq -1$, while for the Sun at $45^{\circ}$ one can apply $\cos \Phi
\simeq -1 / \sqrt{2}$.  The results are shown in Table \ref{table5}.
Comparing these values with those in the last line of Table
\ref{table0} one sees that the ``enhanced'' post-post-Newtonian terms
match the error of the standard post-Newtonian formula.  The
deviation for a grazing ray to the Sun is a few \muas\ and originates
from the post-post-Newtonian terms neglected in (\ref{n_85-better}).

Vector $\ve{n}$ computed using (\ref{n_85-better}) can be denoted as
$\ve{n}^\prime_{\rm pN}$.  The numerical validity of
$\ve{n}^\prime_{\rm pN}$ can be confirmed by direct comparisons of
$\ve{n}^\prime_{\rm pN}$ and vector $\ve{n}$ computed using numerical
integrations of the geodetic equations as discussed in Section
\ref{Sec:numerical-comparison}. For example, the results for Jupiter 
show that the error of $\ve{n}^\prime_{\rm pN}$ does not exceed 0.04 \muas.
The origin of this small deviation is well understood and will be discussed 
elsewhere.

\section{Physical origin of the ``enhanced'' post-post-Newtonian terms}
\label{section-physical-origin}

We have found above the estimates of various terms in the
transformations between units vectors $\ve{\sigma}$, $\ve{n}$, and
$\ve{k}$ characterizing light propagation.  These estimates reveal
that in each transformation ``enhanced'' post-post-Newtonian terms
exist that can become much larger that other ``regular''
post-post-Newtonian terms. In each case the sum of the ``regular''
post-post-Newtonian terms can be estimated as $\frac{\displaystyle
  15}{\displaystyle 4}\,\pi\,\frac{\displaystyle m^2}{\displaystyle
  d^2}$.  The ``enhanced'' terms can be much
larger, being, however, of analytical order $m^2$. In this Section we
clarify the physical origin of the ``enhanced'' terms.

First, let us note that the ``enhanced'' post-post-Newtonian terms in
(\ref{k_to_sigma}), (\ref{n_10}), (\ref{n_60}), and
(\ref{sigma-n-stars}) contain only parameter $\gamma$.  It is clear
that these terms come from the post-Newtonian terms in the metric and
in the equations of motion (parameter $\alpha$ does not appear in
these terms; see Section
\ref{section-equations-with-alpha}). Therefore, their origin is the
formal second-order (post-post-Newtonian) solution of the first-order
(post-Newtonian) equations 
given by the first line of 
(\ref{geodetic-ppN-final-with-alpha}).

Now let us demonstrate that the ``enhanced'' terms result from an inadequate
choice of impact parameters $\ve{d}$ or $\ve{d}_\sigma$ in the standard
post-Newtonian formulas. Indeed, we can 
demonstrate that the ``enhanced'' terms disappear if the light
deflection formulas are expressed through the coordinate-independent
impact parameter $\ve{d}^\prime$ defined by (\ref{d-prime-xdot}).
Eqs.
(\ref{sigma-k-better})--(\ref{sigma-k-S}),
(\ref{n-sigma-better})--(\ref{n-sigma-T}),
(\ref{n_85-better})--(\ref{P-sso}), and
(\ref{sigma-n-stars-simplified})--(\ref{Q-stars})
can be written as
\begin{eqnarray}
\ve{\sigma} &=& \ve{k}\,+\ve{d}^\prime\,S^\prime\,
+{\cal O}\left({m^2\over d^2}\right)+{\cal O}({m^3})
\,,
\label{sigma-k-S-prime}
\\
S^\prime &=& (1+\gamma) {m\over d^{\prime2}} \left(1-{x_{\source}-x_{\obs}\over R}\right),
\label{S-prime}
\\
\ve{n}&=&\ve{\sigma}+\ve{d}^{\;\prime}\,T^{\;\prime} 
+ {\cal O} \left(\frac{m^2}{d^2}\right) + {\cal O} \left(m^3\right)\,,
\label{n-sigma-postNewtonian_10}
\\
T^{\prime} &=& -(1 + \gamma)\,\frac{m}{d^{\;\prime2}}\,
\left(1 + \frac{\ve{k} \cdot \ve{x}}{x}\right)\,,
\label{T-prime}
\\
\ve{n}&=&\ve{k}+\ve{d}^{\;\prime}\,P^{\;\prime} 
+ {\cal O}\left(\frac{m^2}{d^2}\right)\, + {\cal O}\left(m^3\right)\,,
\label{n_85-postNewtonian_10}
\\
P^\prime &=&-(1+\gamma)\,{m\over d^{\prime2}}\,
\left({x_0-x\over R}+{\ve{k}\cdot\ve{x}\over x}\right)\,,
\label{P-sso-prime}
\\
\ve{n} &=& \ve{\sigma}+\ve{d}^{\;\prime}\,Q^{\;\prime} + {\cal O}\left(\frac{m^2}{d_{\sigma}^2}\right) 
+ {\cal O}\left(m^3\right)\,,
\label{sigma-n-stars-postNewtonian_10}
\\
Q^\prime &=& -(1+\gamma)\,{m\over d^{\;\prime2}}\,\left(1+{\ve{\sigma}\cdot\ve{x}\over x}\right)\,,
\label{Q-stars-prime}
\end{eqnarray}
\noindent
respectively. Therefore, in each case the ``enhanced''
post-post-Newtonian terms only correct the post-Newtonian terms
that use inadequate impact parameter.
Let us stress, however, that for practical calculations
(\ref{sigma-k-better})--(\ref{sigma-k-S}),
(\ref{n-sigma-better})--(\ref{n-sigma-T}),
(\ref{n_85-better})--(\ref{P-sso}), and
(\ref{sigma-n-stars-simplified})--(\ref{Q-stars}) are more convenient.

\section{Summary and concluding remarks}
\label{section-conclusion}

In this paper the numerical accuracy of the post-Newtonian and
post-post-Newtonian formulas for light propagation in the parametrized
Schwarzschild field has been investigated. Analytical formulas have
been compared with high-accuracy numerical integrations of the
geodetic equations. In this way we demonstrate that the standard
post-Newtonian formulas for the boundary problem (light propagation
between two given points) cannot be used at the accuracy level of
1~\muas\ for observations performed by an observer situated within the
solar system. The error of the standard formula may attain $\sim$16
\muas. Detailed analysis has shown that the error is of
post-post-Newtonian order ${\cal O}(m^2)$. On the other hand, the
post-post-Newtonian terms are often thought to be of order $m^2/d^2$
and can be estimated to be much smaller than 1 \muas\ in this case.
To clarify this contradiction we have derived and investigated the
explicit analytical post-post-Newtonian solution for the light
propagation. For each individual term in the relevant formulas exact
analytical upper estimates have been found. It turns out that in each
case there exist post-post-Newtonian terms that can become much larger
than the other ones and cannot be estimated as ${\rm const}\times
m^2/d^2$. We call these terms ``enhanced'' post-post-Newtonian terms.
These terms depend only on $\gamma$ and come from the second-order
solution of the post-Newtonian equations of light propagation
(Eq. (39) with $\alpha=0$). For this reason one could argue that the
``enhanced'' post-post-Newtonian terms should not be called
``post-post-Newtonian'', but better ``$m^2$-terms'' or similarly. The
physical origin of the ``enhanced'' terms is discussed in the previous
Section.  The derived analytical solution shows that no ``regular''
post-post-Newtonian terms are relevant for the accuracy of 1 \muas\ in
the conditions of planned astrometric missions (Gaia, SIM, etc.). Most
of the ``regular'' terms come from the post-post-Newtonian terms in
the metric tensor.  It is not the post-Newtonian equation of light
propagation (Eq. (39) with $\alpha=0$) itself, but the standard
analytical way to solve this equation that is responsible for the
numerical error of 16 \muas\ mentioned above.

The compact formulas for the light propagation time and for the
transformations between directions $\ve{\sigma}$, $\ve{n}$ and
$\ve{k}$ have been derived.  The formulas are given by (\ref{tau_30}),
(\ref{sigma-k-better})--(\ref{sigma-k-S}),
(\ref{n-sigma-better})--(\ref{n-sigma-T}),
(\ref{n_85-better})--(\ref{P-sso}), and
(\ref{sigma-n-stars-simplified})--(\ref{Q-stars}).  These formulas
contain only terms (both post-Newtonian and post-post-Newtonian) that
are numerically relevant at the level of 10 cm for the Shapiro delay
and 1 \muas\ for the directions for any observer situated in the solar
system and not observing closer than 3.3 angular radii of the
Sun.

Let us finally note that the post-post-Newtonian term in
(\ref{n_85-better})--(\ref{P-sso}) is closely related to the 
gravitational lens formula. Here we only note that all the formulas
for the boundary problem given in this paper are not valid for $d=0$
($d$ always appear in the denominators of these formulas). On the
other hand, the standard post-Newtonian lens equation successfully
treats this case known as the Einstein ring solution. The relation
between the lens approximation and the standard post-Newtonian
expansion is a different topic which will be considered in a
subsequent paper.

\ack

This work was partially supported by the BMWi grants 50\,QG\,0601 and
50\,QG\,0901 awarded by the Deutsche Zentrum f\"ur Luft- und Raumfahrt
e.V. (DLR).

\end{document}